\documentclass[twocolumn,epjc3]{svjour3}

\RequirePackage[T1]{fontenc}
\smartqed
\usepackage{graphicx}
\graphicspath{{./}}
\usepackage{mathptmx}
\usepackage{flushend}
\usepackage[numbers,sort&compress]{natbib}
\usepackage{amsmath,amssymb}
\usepackage{dcolumn}
\usepackage{bm}
\usepackage{booktabs}
\usepackage{CJKutf8}
\usepackage{tikz,xcolor}
\usepackage{hyperref}
\journalname{Eur. Phys. J. C}

\hypersetup{
  colorlinks=true,
  linkcolor=blue,
  urlcolor=blue,
  citecolor=blue,
  linktoc=all
}

\definecolor{lime}{HTML}{A6CE39}
\DeclareRobustCommand{\orcidicon}{%
  \begin{tikzpicture}
  \draw[lime, fill=lime] (0,0)
  circle [radius=0.16]
  node[white] {{\fontfamily{qag}\selectfont \tiny ID}};
  \draw[white, fill=white] (-0.0625,0.095)
  circle [radius=0.007];
  \end{tikzpicture}
  \hspace{-2mm}
}

\foreach \x in {A, ..., Z}{%
  \expandafter\xdef\csname orcid\x\endcsname{\noexpand\href{https://orcid.org/\csname orcidauthor\x\endcsname}{\noexpand\orcidicon}}
}

\begin{document}
\sloppy
\begin{CJK*}{UTF8}{gbsn}

\title{From Matched Black-Hole Scalar Charges to Gravitational-Wave Inspirals in a Massive Scalar--Gauss--Bonnet Effective Theory}
\author{Jie Wu\thanksref{addr1,addr2}\orcidA{}
        \and Mengfei Sun\thanksref{addr1,addr2}\orcidB{}
        \and Jin Li\thanksref{addr1,addr2,addr3,e}\orcidC{}}
\thankstext{e}{e-mail: \href{mailto:cqujinli1983@cqu.edu.cn}{cqujinli1983@cqu.edu.cn}}
\institute{College of Physics, Chongqing University, Chongqing 401331, China\label{addr1}\hypertarget{addr1}{}
        \and Department of Physics and Chongqing Key Laboratory for Strongly Coupled Physics, Chongqing University, Chongqing 401331, China\label{addr2}\hypertarget{addr2}{}
        \and Institute of Advanced Interdisciplinary Studies, Chongqing University, Chongqing 401331, China\label{addr3}\hypertarget{addr3}{}}
\maketitle
\begin{abstract}
Massive scalar fields introduce a finite interaction range in scalar--Gauss--Bonnet (sGB) gravity and can modify the gravitational waves (GWs) emitted by binary black holes (BBHs). We develop a matched description that relates the curvature-sourced scalar field of an isolated black hole (BH) to the scalar charge entering the effective binary dynamics. The massive field contains a distributed curvature-forced contribution and a homogeneous Yukawa sector. By matching the near- and far-zone solutions, we determine the local static coefficient and derive its finite conversion to a renormalized worldline coupling. The static coefficient has no linear scalar-mass correction, and its first correction is quadratic. We use the leading matched charges in the leading post-Newtonian dynamics and GW phasing of nonspinning, quasicircular BBHs while retaining the finite Yukawa range and massive-radiation kinematics. The scalar mass modifies the conservative dynamics, produces distinct dipole and quadrupole radiation thresholds, and changes the accumulated inspiral phase. We also construct the associated breathing and longitudinal tidal responses and treat dispersive massive propagation separately. These results connect the finite-mass BH scalar response to source and propagation observables for BBH inspirals.
\end{abstract}
\end{CJK*}

\section{Introduction}

Gravitational waves (GWs) from compact binaries probe gravity in the dynamical strong-field regime. The direct detection of GW150914 established this observational channel~\cite{Abbott2016GW150914}. Signals from binary black holes (BBHs) provide a direct setting for tests of General Relativity (GR)~\cite{Yunes2025Tests}. The LIGO--Virgo--KAGRA (LVK) catalog from the fourth observing run has enlarged the sample of compact-binary signals~\cite{LVK2025GWTC4}. Recent LVK catalog-based tests find no statistically significant departure from GR~\cite{LVK2026GWTC5GR}. These observations motivate calculations that relate compact-object structure to the dynamics and radiation of binaries~\cite{Berti2015Tests}. Long-duration inspirals observed by future space-based detectors offer a complementary regime, where resolving small cumulative phase corrections requires accurate waveform models~\cite{my_paper_PN}.

Many extensions of GR contain one or more scalar degrees of freedom~\cite{Will2014Confrontation,Damour1996Pulsar}. A scalar field can endow a compact object with a response that depends on its internal gravitational binding~\cite{Eardley1975Scalar}. Such responses can become nonperturbative in strong gravity~\cite{Damour1993StrongField}. They modify the conservative motion and generate dipole radiation in compact binaries~\cite{Mirshekari2013Motion,Lang2015ScalarWaves}. Waveform and phasing calculations in scalar-tensor gravity determine how compact-object properties enter the radiation and phase evolution~\cite{Sennett2016Waveforms}. Effective field theory (EFT) organizes the associated worldline couplings, near-zone dynamics, and radiation~\cite{Kuntz2019EFT}. Post-Newtonian (PN) calculations in quadratic gravity show the importance of matching compact objects to appropriate effective sources~\cite{Yagi2012Quadratic}.
The scalar field also produces nontensorial GW polarizations that can be tested with ground- and space-based detector networks~\cite{my_paper_polarization}. The distinction between a generated scalar field and its physical metric response also appears in scalar waveform studies~\cite{Zosso2026ScalarMemory}.

Scalar--Gauss--Bonnet (sGB) gravity provides a geometric extension in which a scalar field couples directly to the Gauss--Bonnet (GB) invariant. Early higher-curvature constructions established black-hole (BH) solutions with a nontrivial scalar field~\cite{Kanti1996DilatonicBH}. The linear sGB coupling gives a curvature-sourced BH response~\cite{Sotiriou2014Hair,Sotiriou2014Example}. Curvature-induced scalarization gives additional branches of BH solutions~\cite{Doneva2018Scalarization,Silva2018Scalarization}. Rotating solutions and perturbative spectra have extended the strong-field description of these BHs~\cite{Ayzenberg2014RotatingBH,BlazquezSalcedo2016Perturbations}. Perturbative and nonlinear studies address their stability and evolution in isolated and binary configurations~\cite{Ripley2020Dynamics,Silva2019Stability,East2021Evolution,Lara2024Scalarization}.

The scalar mass introduces a second physical scale. Scalar exchange then has a finite Yukawa range~\cite{Alsing2012MassiveBD}. A radiative harmonic propagates only when its frequency exceeds the scalar mass, and the resulting massive propagation is dispersive~\cite{Diedrichs2024MassivePhase}. Massive sGB BH solutions show that the scalar mass can reshape the local response~\cite{Doneva2019MassiveBH,Peng2020MassiveScalarization}. Small-coupling calculations of massive scalar clouds determine the leading scalar response on a Schwarzschild background and its higher-order metric backreaction~\cite{vanGemeren2024MassiveClouds}.

In massless sGB gravity, Juli\'e and Berti connected BH skeletonization to a PN description of binary dynamics~\cite{Julie2019Dynamics}, and Shiralilou \emph{et al.} computed gravitational and scalar waves for inspirals~\cite{Shiralilou2022Waves}. BH sensitivities give a systematic connection between isolated solutions and point-particle descriptions~\cite{Julie2022Sensitivities}. Binary-BH evolution and merger have also been studied in sGB gravity~\cite{Witek2019Mergers,East2021Scalarization,Corman2023NonlinearBBH}. Quasistationary initial data have made the binary scalar profile accessible before an evolution begins~\cite{Nee2025InitialData}. Recent simulations also examine dynamical hair growth during the inspiral~\cite{Capuano2026HairGrowth}.

For a massive scalar, van Gemeren \emph{et al.} derived compact-binary dynamics through 1PN order~\cite{vanGemeren2026Dynamics}. That analysis also emphasized that relating the exponentially decaying scalar response of a massive BH to worldline parameters requires a nontrivial matching prescription. Their illustrative binary calculations use the massless charge values, while the matching to massive BH solutions is left for future work~\cite{vanGemeren2026Dynamics}. The massive response contains both a distributed forced sector and a homogeneous Yukawa sector, so a far-zone scalar amplitude alone does not specify the binary effective scalar charge.

In this work, we match the curvature-sourced BH response between the near and far zones through their common overlap region and obtain the local static matching coefficient. The linear finite-mass correction vanishes, and the first nonzero correction is quadratic, in agreement with an independent perturbative numerical reconstruction. We derive the finite conversion from this static coefficient to a renormalized worldline coupling and use the leading matched charges in the leading-PN binary dynamics, scalar radiation, and inspiral phase while retaining the finite Yukawa range. The scalar field is then mapped to the breathing and longitudinal tidal responses, while local massive propagation is treated separately. This sequence connects the isolated-BH solution directly to BBH inspiral observables.

The paper is organized as follows. In Sec.~\ref{sec:bh-scalar-matching}, we introduce the effective model and approximation hierarchy, derive the BH scalar matching, and present numerical checks. Section~\ref{sec:binary-dynamics} presents the binary dynamics, scalar radiation, inspiral evolution, and Fourier phase. The scalar waveform and massive propagation are described in Sec.~\ref{sec:scalar-waveform}. Sec.~\ref{sec:discussion} discusses the implications and scope of the results, the principal limitations, and future directions. \ref{app:scalar-matching} gives technical details of the scalar matching, and \ref{app:worldline-radiation} derives the worldline normalization and leading scalar radiation. Throughout this paper, we use geometrized units $c=G=1$.

\section{Black-hole scalar matching}
\label{sec:bh-scalar-matching}

\subsection{Effective model and approximation hierarchy}

\begin{figure*}[!t]
\centering
\includegraphics[width=0.98\textwidth]{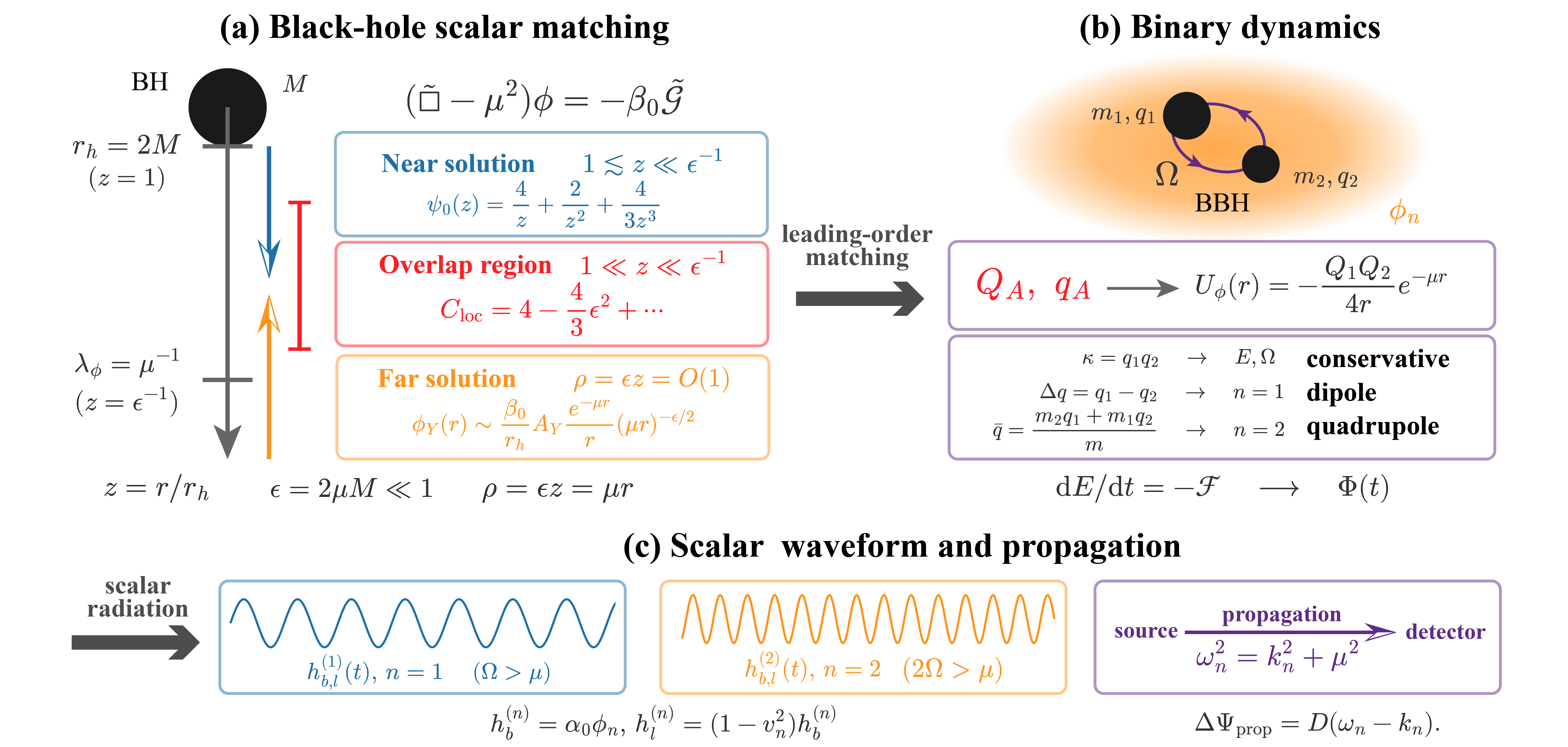}
\caption{Schematic overview of the calculation. (a) BH scalar matching connects the near and far scalar solutions through their overlap region, yielding the static coefficient $C_{\rm loc}$ (Sec.~\ref{sec:bh-scalar-matching}). (b) Binary dynamics uses the leading matched charges in finite-range interactions, with $\kappa$ controlling the retained conservative scalar correction and $\Delta q$ and $\bar q$ controlling the $n=1$ and $n=2$ scalar-radiation sectors, while energy balance determines the orbital phase (Sec.~\ref{sec:binary-dynamics}). (c) The $n=1$ and $n=2$ scalar harmonics generate linked breathing and longitudinal tidal responses, and massive propagation is dispersive (Sec.~\ref{sec:scalar-waveform}). The schematic wave trains indicate harmonic content rather than the relative physical amplitudes of $h_b$ and $h_l$.}
\label{fig:overview}
\end{figure*}

The massive sGB effective theory introduces a massive scalar degree of freedom and the leading curvature coupling responsible for the scalar response~\cite{Doneva2019MassiveBH}. We calculate the strong-field scalar response that supplies the compact-object input to the binary description. The scalar mass is treated as an effective extension of the sGB sector, with the leading conformal matter coupling included. The calculation is performed at leading order in the sGB coupling on a GR BH background. The sequence from the isolated-BH response to the binary observables is summarized in Fig.~\ref{fig:overview}.

The theory is formulated in the Einstein frame through the action~\cite{Doneva2019MassiveBH,Evstafyeva2023ScalarPolarization}
\begin{equation}
\label{eq:effective-action}
\begin{aligned}
S={}&\frac{1}{16\pi}\int \mathrm{d}^4x\sqrt{-\tilde g}\left[\tilde R-\frac{1}{2}(\tilde\nabla\phi)^2-\frac{1}{2}\mu^2\phi^2+\beta_0\phi\tilde{\mathcal G}\right] \\
&+S_m[\psi_m,g_{\mu\nu}].
\end{aligned}
\end{equation}
The metric $\tilde g_{\mu\nu}$ defines the Einstein-frame geometry, and $\tilde R$ is its Ricci scalar. The Einstein-frame scalar field $\phi$ has mass $\mu$. The coupling $\beta_0$ has dimensions of length squared and controls the interaction between $\phi$ and the GB invariant, while $\psi_m$ collectively denotes the matter fields. The GB invariant $\tilde{\mathcal G}$ is~\cite{Sotiriou2014Example}
\begin{equation}
\label{eq:gauss-bonnet-invariant}
\tilde{\mathcal G}=\tilde R_{\mu\nu\rho\sigma}\tilde R^{\mu\nu\rho\sigma}
-4\tilde R_{\mu\nu}\tilde R^{\mu\nu}+\tilde R^2,
\end{equation}
and every curvature quantity in Eqs.~(\ref{eq:effective-action}) and (\ref{eq:gauss-bonnet-invariant}) is constructed from $\tilde g_{\mu\nu}$.

Matter couples to the physical (Jordan-frame) metric through the linearized conformal relation~\cite{Damour1996Pulsar,Evstafyeva2023ScalarPolarization}
\begin{equation}
\label{eq:physical-metric-relation}
g_{\mu\nu}=(1+\alpha_0\phi)\tilde g_{\mu\nu},
\end{equation}
where $\alpha_0$ characterizes the leading-order conformal coupling between the scalar field and the physical metric. The coupling $\beta_0$ controls the curvature-sourced scalar response and the matched scalar charges, whereas $\alpha_0$ converts the canonical scalar field into a physical metric response. At the retained order, this coupling affects the mapping to the physical metric while leaving the vacuum isolated-BH matching and the associated charge $Q_A$ unchanged. A tilde marks Einstein-frame metric and curvature quantities, while $g_{\mu\nu}$ denotes the physical metric entering the matter action.

In the vacuum exterior of the isolated BH, variation with respect to $\phi$ gives the sourced massive scalar equation~\cite{Doneva2019MassiveBH}
\begin{equation}
\label{eq:scalar-equation}
(\tilde\Box-\mu^2)\phi=-\beta_0\tilde{\mathcal G},
\end{equation}
where $\tilde\Box\equiv\tilde g^{\mu\nu}\tilde\nabla_\mu\tilde\nabla_\nu$ is the covariant d'Alembertian. The leading scalar field is linear in \mbox{$\displaystyle \beta_0$}, while its stress tensor and the GB contribution to the Einstein-frame metric equation start at quadratic order in \mbox{$\displaystyle \beta_0$}~\cite{vanGemeren2024MassiveClouds}. We therefore compute the isolated scalar response on a fixed GR Schwarzschild background. Keeping the scalar-mass expansion through second order does not change this coupling-order counting. This step determines the curvature-sourced scalar field before binary interactions are introduced.

For an isolated Schwarzschild BH of mass $M$, the local small parameter is $\epsilon\equiv\mu r_h\ll1$, where $r_h=2M$. Weak scalar coupling and small $\epsilon$ are independent assumptions. The former permits the scalar response to be calculated on a fixed GR background, while the latter controls the local near-to-far matching. The local matching expands in $\epsilon$, whereas the binary analysis retains the finite Yukawa range and massive-radiation thresholds at finite $\mu r$~\cite{Alsing2012MassiveBD,Diedrichs2024MassivePhase}. Far-zone Yukawa behavior alone does not define the effective scalar charge used in binary dynamics. This distinction motivates the local matching prescription developed in Sec.~\ref{sec:scalar-matching}.

\subsection{Scalar matching and local charge}
\label{sec:scalar-matching}

For a static, spherically symmetric scalar field $\phi(r)$, the background line element is
\begin{equation}
\mathrm{d}\tilde{s}^{2}=-f(r)\mathrm{d}t^{2}+f(r)^{-1}\mathrm{d}r^{2}+r^{2}\left(\mathrm{d}\vartheta^{2}+\sin^{2}\vartheta\,\mathrm{d}\varphi^{2}\right),
\end{equation}
where $f(r)=1-2M/r$, $M$ is the isolated BH mass, and $\vartheta$ and $\varphi$ are the usual angular coordinates. The Schwarzschild GB invariant is $\tilde{\mathcal G}=48M^2/r^6$~\cite{Sotiriou2014Example,Doneva2019MassiveBH}. The radial relation following from Eq.~(\ref{eq:scalar-equation}) takes the form
\begin{equation}
\frac{1}{r^2}\frac{\mathrm{d}}{\mathrm{d}r}\left[r^2 f(r)\frac{\mathrm{d}\phi}{\mathrm{d}r}\right]-\mu^2\phi=-\frac{48\beta_0M^2}{r^6}.
\end{equation}
The radial equation is inhomogeneous at every finite radius because the curvature source is distributed throughout the exterior spacetime. The horizon expansion and coefficient extraction are summarized in \ref{app:scalar-matching}.

Using the horizon radius $r_h=2M$, we introduce $z=r/r_h$, $\epsilon=\mu r_h=2\mu M$, and $\psi(z)=r_h^2\phi(r)/\beta_0$. The parameter $\epsilon\ll1$ compares the horizon scale with the scalar Compton wavelength. With these variables, the radial equation takes the form
\begin{equation}
\label{eq:dimensionless-radial-equation}
\frac{\mathrm{d}}{\mathrm{d}z}\left[z(z-1)\frac{\mathrm{d}\psi}{\mathrm{d}z}\right]-\epsilon^2 z^2\psi=-\frac{12}{z^4}.
\end{equation}
Horizon regularity selects the branch that is regular on the future horizon. At $\epsilon=0$, this branch approaches the decaying massless solution
\begin{equation}
\psi_0(z)=\frac{4}{z}+\frac{2}{z^2}+\frac{4}{3z^3}.
\end{equation}
Its leading large-$z$ term is the local $1/z$ monopole used in the overlap matching. We expand the local response in small $\epsilon$ while keeping the massive far-zone behavior explicit.

The massive solution is not a pure Yukawa field. The nonvanishing GB source generates a curvature-induced particular solution. Its leading far behavior follows from the radial equation
\begin{equation}
\label{eq:forced-tail}
\phi_{\rm forced}(r)\sim\frac{48\beta_0M^2}{\mu^2r^6}.
\end{equation}
This asymptotic behavior applies at large radius in the regime $\mu r\gg1$, where the massive term balances the curvature source. The forced contribution must be separated before extracting the homogeneous Yukawa amplitude.

For the far-zone amplitude extraction, \mbox{$\displaystyle \psi_{\rm p}(z)$} denotes the particular solution selected by the finite-boundary subtraction specified in Sec.~\ref{sec:numerical-validation}. The complete solution is decomposed into this particular solution and a homogeneous remainder. The source-subtracted field is
\begin{equation}
\psi_{\rm hom}(z)=\psi(z)-\psi_{\rm p}(z).
\end{equation}
The field $\psi_{\rm hom}$ satisfies the corresponding homogeneous radial equation. This subtraction fixes the homogeneous remainder used for the numerical amplitude extraction. Redistributing a decaying homogeneous solution between $\psi_{\rm p}$ and $\psi_{\rm hom}$ leaves the total field and, at fixed local normalization, $C_{\rm loc}$ unchanged.

The homogeneous equation has decaying and growing large-$z$ basis functions
\begin{equation}
\label{eq:distorted-yukawa-basis}
u_{\rm dec}(z)\sim \mathrm{e}^{-\epsilon z}z^{-1-\epsilon/2},\qquad
u_{\rm grow}(z)\sim \mathrm{e}^{+\epsilon z}z^{-1+\epsilon/2}.
\end{equation}
Writing $\psi_{\rm hom}=C_{\rm dec}u_{\rm dec}+C_{\rm grow}u_{\rm grow}$ defines $C_{\rm dec}$ as the normalization of the Schwarzschild-distorted decaying homogeneous mode. The asymptotically decaying solution requires $C_{\rm grow}=0$. The factors $z^{\mp\epsilon/2}$ encode weak-curvature propagation distortion, as illustrated by the matching chain in Fig.~\ref{fig:overview}(a).

The local monopole is fixed in the overlap of the near and far descriptions. The near zone satisfies $1\lesssim z\ll\epsilon^{-1}$, the far zone uses $\rho=\epsilon z=\mathcal O(1)$, and the common domain is $1\ll z\ll\epsilon^{-1}$. The source-subtracted homogeneous scalar field is denoted by $\phi_Y(r)$. The dimensionless far-zone Yukawa amplitude $A_Y$ is defined by
\begin{equation}
\phi_Y(r)\sim\frac{\beta_0}{r_h}A_Y\frac{\mathrm{e}^{-\mu r}}{r}(\mu r)^{-\epsilon/2}.
\end{equation}
$A_Y$ normalizes the far-zone Yukawa field after the Schwarzschild basis conversion. The factor $(\mu r)^{-\epsilon/2}$ gives the leading curvature distortion of the decaying far-zone mode. Converting this far-zone normalization to the local monopole coefficient requires the prescribed normalization and overlap matching~\cite{vanGemeren2026Dynamics}. The basis conversion, whose asymptotic matching is detailed in \ref{app:scalar-matching}, gives
\begin{equation}
\label{eq:ay-cdec-relation}
A_Y=C_{\rm dec}\epsilon^{\epsilon/2}.
\end{equation}

We extract $C_{\rm loc}$ order by order in $\epsilon$ from the complete near-zone field in Schwarzschild areal coordinates, separating the logarithmic $1/z$ terms before reading off the coefficient of $1/z$. The complete near-to-far matching calculation gives
\begin{equation}
\label{eq:local-coefficient-massless-limit}
C_{\rm loc}=4-\frac{4}{3}\epsilon^2+\mathcal O\!\left(\epsilon^3\log^p\epsilon\right).
\end{equation}
The linear correction to this static near-zone coefficient cancels, and its first nonvanishing correction is the explicit quadratic term $-4\epsilon^2/3$. The symbol $p$ denotes an unspecified finite logarithmic power. The remainder records the asymptotic order of the local matching expansion.

The overlap region used above is the common domain of the near- and far-zone asymptotic expansions. The binary EFT uses a buffer region outside the isolated-BH strong-field zone and inside the binary separation scale, involving a different hierarchy of scales. Matching the BH and point-source fields in the same coordinates relates the static coefficient to the worldline coupling~\cite{Julie2019Dynamics,Ivanov2023Matching}. At second order, this comparison includes the curvature correction to the point-source field. We define the worldline coupling using dimensional regularization and minimal subtraction, removing only the poles. Its source normalization is the one derived in \ref{app:worldline-radiation}. The static matching gives
\begin{equation}
\label{eq:static-worldline-conversion}
\mbox{$\displaystyle \begin{aligned}
Q(\ell)={}&\frac{\beta_0}{r_h}C_{\rm loc}\biggl\{1+\epsilon^2\biggl[\frac43\\
&+\frac{\gamma_{\rm E}+\log(4\pi)}{6}
+\frac13\log\frac{r_h}{\ell}\biggr]\biggr\}+\cdots.
\end{aligned}$}
\end{equation}
The length $\ell$ is the subtraction scale, and $\gamma_{\rm E}$ is the Euler--Mascheroni constant. The finite conversion follows from the unit-source calculation in \ref{app:scalar-matching}. Through the retained order, the scale dependence of the coupling cancels against that of the monopole sector of the point-source field.

For a BBH, the mass of each component is denoted by $m_A$, where $A=1,2$. The corresponding local parameter is $\epsilon_A\equiv2\mu m_A$. The leading matched local scalar charge and its dimensionless form, used in the binary application below, are
\begin{equation}
\label{eq:leading-matched-charges}
\mbox{$\displaystyle Q_A=\frac{2\beta_0}{m_A},\qquad q_A=\frac{Q_A}{2m_A}=\frac{\beta_0}{m_A^2}.$}
\end{equation}
For a massive scalar, this local worldline charge is not a conserved asymptotic charge defined by a Gauss law at spatial infinity.

\subsection[Numerical validation]{Numerical validation}
\label{sec:numerical-validation}

\begin{figure}[!ht]
\centering
\includegraphics[width=0.96\columnwidth]{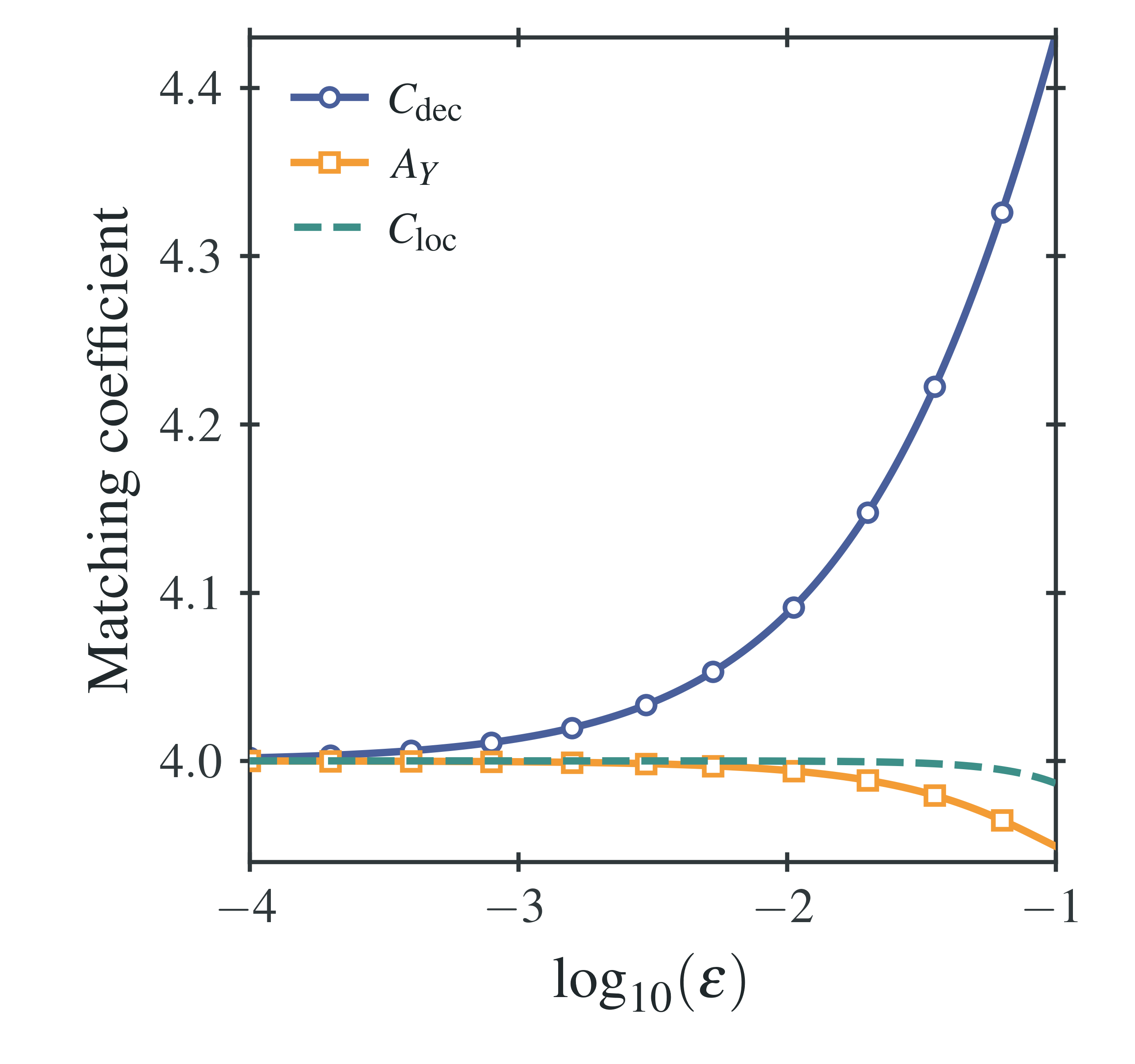}
\caption{Near-to-far matching of the scalar response in the small-$\epsilon$ domain. Blue circles show the numerical Schwarzschild-distorted mode coefficient $C_{\rm dec}$, and orange squares show the converted far-zone Yukawa amplitude $A_Y$. The teal dashed curve is the analytic local coefficient $C_{\rm loc}$ from Eq.~(\ref{eq:local-coefficient-massless-limit}).}
\label{fig:matching-coefficients}
\end{figure}

Separate numerical checks test the radial solution, the quadratic static coefficient, and the Green-identity implementation. We solve the isolated Schwarzschild radial problem by linear shooting from the outer boundary toward the horizon on a truncated radial domain. The value and radial derivative of the numerical forced branch at the outer boundary are set by the truncated inverse-power forced series. A normalized decaying homogeneous branch is initialized with the Schwarzschild-corrected outer series. The two branches are integrated inward separately. The horizon regularity relation in Eq.~(\ref{eq:horizon-regularity-relation}) then fixes $C_{\rm dec}$ algebraically. We project the source-subtracted homogeneous field onto the decaying and growing basis functions and use the extracted $C_{\rm dec}$ values in Fig.~\ref{fig:matching-coefficients}. The residual growing component is used only as a numerical contamination diagnostic.

An independent collocation boundary-value problem (BVP) imposes the horizon regularity and Schwarzschild-corrected outer conditions directly and cross-checks the shooting result. The shooting and collocation solutions give consistent radial profiles in the overlap and far-projection regions. At $\epsilon=0.01,0.02,0.05$, the collocation solutions reach maximum root-mean-square residuals between $1.33\times10^{-9}$ and $2.00\times10^{-9}$. Using the same far-basis projection, the relative differences in $C_{\rm dec}$ between shooting and collocation remain below $1.51\times10^{-8}$.

We independently reconstruct the perturbative $\mathcal O(\epsilon^2)$ radial problem and fit the local $1/z$ coefficient freely to test the analytic value. The convergence of this reconstruction is reported in Table~\ref{tab:quadratic-reconstruction-convergence}. A finite-radius projection of the complete numerical field checks the Green-identity implementation with the analytically fixed normalization of the regular homogeneous solution. The results are reported in Table~\ref{tab:worldline-projection}.

The matching relation is illustrated in Fig.~\ref{fig:matching-coefficients}. All three coefficients approach the common massless value $4$ as $\epsilon\to0$. Their finite-$\epsilon$ separation reflects the distinct normalization regions. The pronounced variation of $C_{\rm dec}$ is an effect of the curvature-distorted far-zone basis. The static curve shows the truncated near-zone coefficient, while the binary calculation uses the leading values in Eq.~(\ref{eq:leading-matched-charges}).

\section{Binary dynamics}
\label{sec:binary-dynamics}

\subsection{Conservative dynamics and scalar radiation}
\label{sec:conservative-dynamics}

We use the leading matched charges in the binary dynamics and radiation, retaining the finite Yukawa range and massive-radiation kinematics. The second-order static conversion is kept in the isolated matching analysis. A skeletonized description represents the matched BH response through an Einstein-frame scalar-dependent mass function~\cite{Julie2019Dynamics,Julie2022Sensitivities,Kuntz2019EFT}. Around a vanishing external scalar field,
\begin{equation}
m_A(\phi;\mu)=m_A-\frac{Q_A}{4}\phi+\mathcal O(\phi^2).
\end{equation}
The linear coefficient is fixed by $Q_A$, with the normalization derived in \ref{app:worldline-radiation}. In the present normalization, $q_A=-2\,\partial_\phi\ln m_A|_{\phi=0}$. The field $\phi$ in this mass function denotes the regular external scalar entering the matched worldline coupling. Higher scalar response and derivative worldline operators lie outside the retained truncation.

For the binary, $m=m_1+m_2$ denotes the total mass and $\eta=m_1m_2/m^2$ the symmetric mass ratio. The leading matched charges give the combinations
\begin{equation}
\label{eq:finite-mass-charge-combinations}
\begin{aligned}
\Delta q&=q_1-q_2
=\beta_0\left(\frac{1}{m_1^2}-\frac{1}{m_2^2}\right),\\
\bar q&=\frac{m_2q_1+m_1q_2}{m}
=\mbox{$\displaystyle \frac{\beta_0}{m}\left(\frac{m_2}{m_1^2}+\frac{m_1}{m_2^2}\right)$},\\
\kappa&=q_1q_2
=\mbox{$\displaystyle \frac{\beta_0^2}{m_1^2m_2^2}$}.
\end{aligned}
\end{equation}
The charge difference controls scalar dipole radiation, the weighted charge controls scalar quadrupole radiation, and the product controls the conservative interaction.

We use a $30\,\mathrm{M}_\odot+20\,\mathrm{M}_\odot$ binary as the benchmark below, with $\beta_0=0.02(20\,\mathrm{M}_\odot)^2$. The leading charges give \mbox{$\displaystyle q_1\simeq8.9\times10^{-3}$}, \mbox{$\displaystyle q_2=2.0\times10^{-2}$}, and \mbox{$\displaystyle \kappa\simeq1.8\times10^{-4}$}. Across $0\leq\epsilon_1\leq0.1$, the second component satisfies $\epsilon_2\leq0.067$.

The matched monopoles generate a finite-range scalar exchange in the binary weak zone. Let $\mathbf z_A$ denote the position of body $A$, with $\mathbf r=\mathbf z_1-\mathbf z_2$ and $r=|\mathbf r|$. The exchange potential is
\begin{equation}
\label{eq:yukawa-potential}
U_\phi(r)=-\frac{Q_1Q_2}{4r}\mathrm{e}^{-\mu r}
=-\frac{m_1m_2}{r}\kappa\mathrm{e}^{-\mu r}.
\end{equation}
\ref{app:worldline-radiation} derives the source normalization and attractive sign. Combining the Newtonian gravitational interaction with the Yukawa exchange in Eq.~(\ref{eq:yukawa-potential}) gives the retained relative Lagrangian, with $\mathbf v=\mathrm{d}\mathbf r/\mathrm{d}t$ and $v=|\mathbf v|$,
\begin{equation}
\label{eq:relative-lagrangian}
L=\frac{1}{2}\eta m v^2+\frac{\eta m^2}{r}
\left[1+\kappa\mathrm{e}^{-\mu r}\right],
\end{equation}
with all scalar corrections truncated at $\mathcal O(q_A^2)$. The distributed weak-zone GB contribution is PN suppressed relative to the matched monopole sector. In the binary weak zone, the curvature invariant scales as $\tilde{\mathcal G}_{\rm weak}\sim m^2/r^6$, so the corresponding distributed scalar field scales as $\phi_{\rm GB}^{\rm weak}\sim\beta_0m^2/r^4$. The matched monopole instead scales as $\phi_Q\sim Q/r\sim\beta_0/(mr)$. Their ratio therefore scales as
\begin{equation}
\label{eq:weak-zone-gb-suppression}
\frac{\phi_{\rm GB}^{\rm weak}}{\phi_Q}
\sim
\left(\frac{m}{r}\right)^3
\sim v^6,
\end{equation}
which is a relative 3PN suppression in the weak-coupling counting. The distributed weak-zone source is therefore omitted consistently at the leading point-source order retained here~\cite{Julie2019Dynamics}.

\begin{figure}[!ht]
\includegraphics[width=\columnwidth]{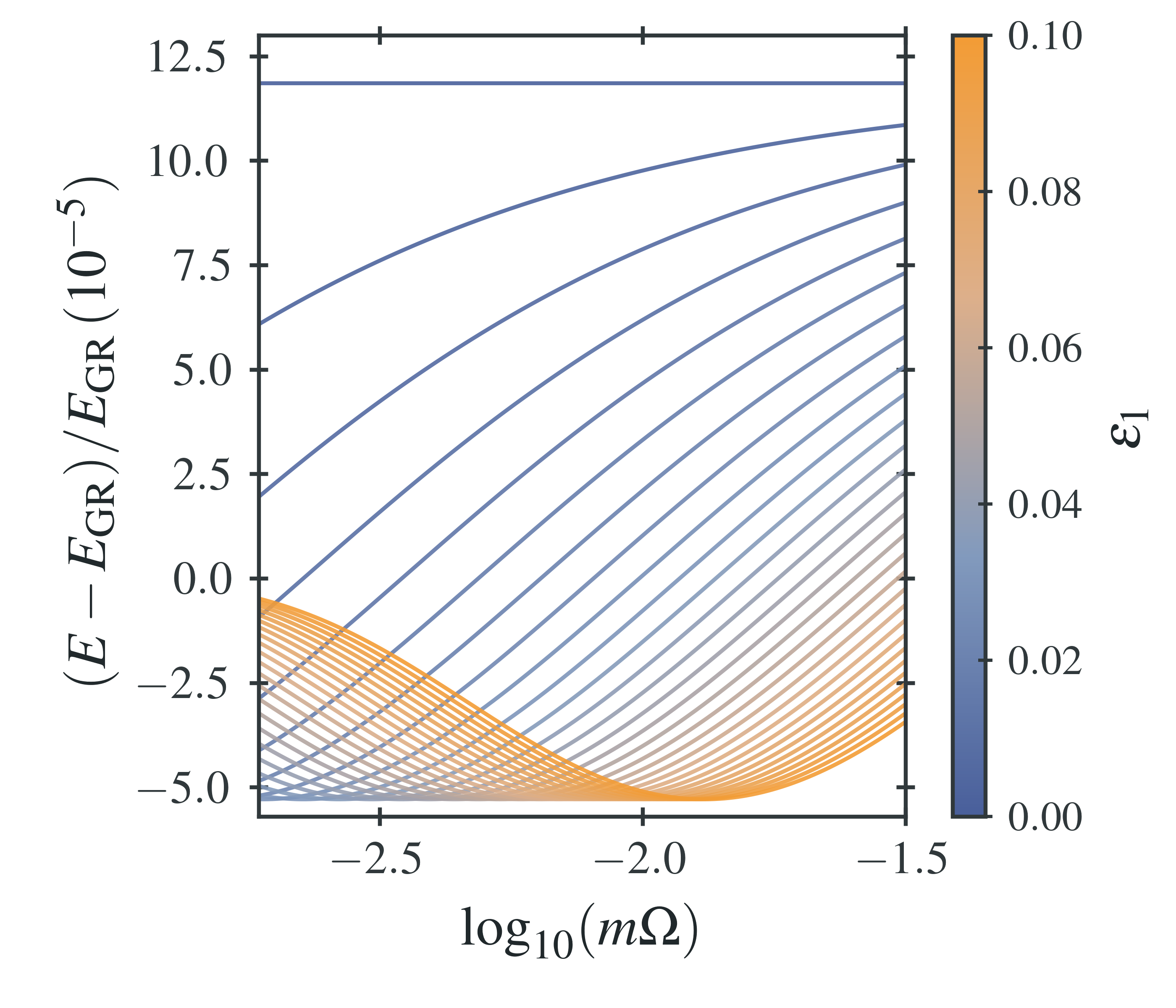}
\caption{Scalar correction to the circular binding energy relative to GR for $m_1=30\,\mathrm{M}_\odot$, $m_2=20\,\mathrm{M}_\odot$, and $\beta_0=0.02(20\,\mathrm{M}_\odot)^2$ in geometrized units. The ordinate shows the fractional deviation from the GR binding energy in units of $10^{-5}$ as a function of $m\Omega$. The color scale samples $0\leq\epsilon_1\leq0.1$, making the continuous dependence on the scalar Compton scale explicit.}
\label{fig:conservative-dynamics}
\end{figure}

Applying the circular-orbit condition to the relative Lagrangian in Eq.~(\ref{eq:relative-lagrangian}) yields the modified Kepler relation, where $\Omega$ is the orbital angular frequency,
\begin{equation}
\label{eq:modified-kepler-relation}
\Omega^2=\frac{m}{r^3}\left[1+\kappa(1+\mu r)\mathrm{e}^{-\mu r}\right].
\end{equation}
The corresponding conserved orbital energy is
\begin{equation}
E(r)=-\frac{\eta m^2}{2r}
\left[1+\kappa(1-\mu r)\mathrm{e}^{-\mu r}\right].
\end{equation}
Defining $x=(m\Omega)^{2/3}$ and $y=\mu m/x$, eliminating $r$ with the modified circular relation, and expanding consistently only in $\kappa$ while leaving $\mathrm{e}^{-y}$ unexpanded yields
\begin{equation}
\label{eq:binding-energy-frequency}
E(\Omega)=-\frac{\eta mx}{2}
\left[1+\frac{2\kappa}{3}(1-2y)\mathrm{e}^{-y}\right].
\end{equation}
The binary-dynamics calculations use $0.015\leq x\leq0.1$. The local matching uses $\epsilon_A\ll1$, whereas the binary-scale variable $y=\mu m/x\simeq\mu r$ is retained in the full Yukawa factor. At the largest scalar mass shown, $y$ decreases from approximately $5.6$ to $0.83$ across the plotted inspiral interval, so the examples probe finite-range suppression with the leading matched charges. The leading-order GR reference is $E_{\rm GR}=-\eta mx/2$, which normalizes the relative energy correction shown in Fig.~\ref{fig:conservative-dynamics}.

At fixed orbital frequency, the small-mass expansion of the binding energy contains a term linear in the scalar mass that is independent of the orbital frequency. It therefore drops out of $\mathrm{d}E/\mathrm{d}x$. Away from radiation thresholds, the source-domain finite-mass correction consequently begins quadratically in the scalar mass at the retained leading-PN order.

The curves in Fig.~\ref{fig:conservative-dynamics} quantify the scalar correction to the circular binding energy relative to GR for different scalar masses. At fixed $x$, a larger $\epsilon_1$ corresponds to a larger scalar mass and a shorter Compton range, so the Yukawa exchange is more strongly suppressed. At fixed $\epsilon_1$, the inspiral toward larger $x$ lowers $y=\mu m/x$ and weakens this range suppression. The signed energy difference follows the finite-range kernel, whose factor $(1-2y)\mathrm e^{-y}$ is nonmonotonic even at fixed charge. The comparison retains the full finite-range factor in $E(\Omega)$ at the present leading-PN order.

\begin{figure}[!ht]
\includegraphics[width=\columnwidth]{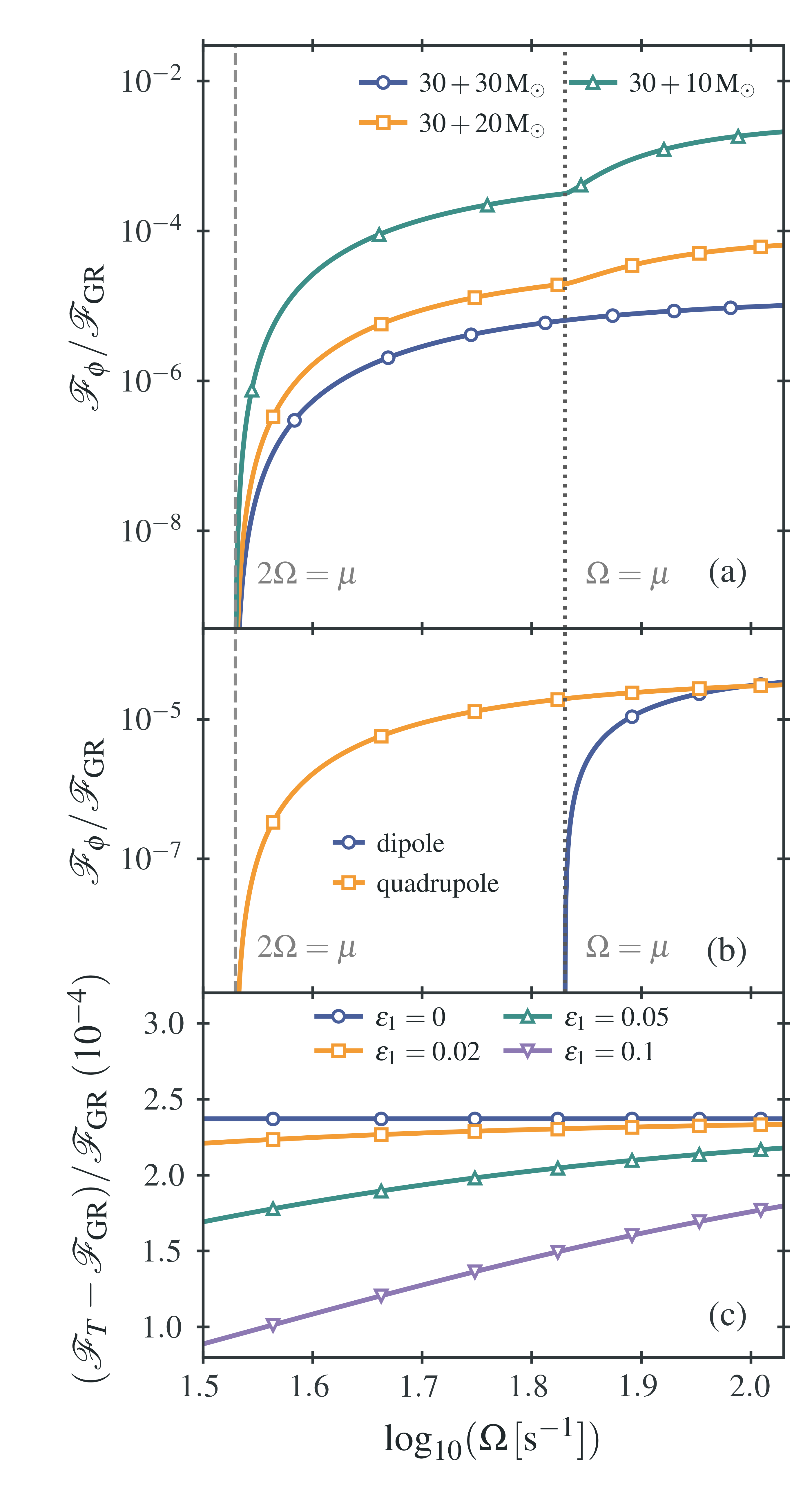}
\caption{Energy-flux corrections in the massive sGB effective theory. (a) Total scalar flux normalized by $\mathcal F_{\rm GR}$, $(\mathcal F_\phi^{\rm dip}+\mathcal F_\phi^{\rm quad})/\mathcal F_{\rm GR}$, for the displayed binary mass combinations at $\epsilon_1=0.020$ and fixed $\beta_0=0.02(20\,\mathrm{M}_\odot)^2$. (b) Scalar dipole and quadrupole fluxes, each normalized by $\mathcal F_{\rm GR}$, for the benchmark and coupling of Fig.~\ref{fig:conservative-dynamics} at \mbox{$\displaystyle \epsilon_1=0.020$}. (c) Tensor flux compared with the GR flux for the same benchmark and coupling at $\epsilon_1=0$, $0.02$, $0.05$, and $0.1$, with the relative difference shown in units of $10^{-4}$. Gray dashed and dotted lines in (a) and (b) mark the $2\Omega=\mu$ and $\Omega=\mu$ radiation thresholds, respectively.}
\label{fig:energy-flux}
\end{figure}

The dissipative sector is determined by the radiative scalar source moments. Using the charge combinations defined in Eq.~(\ref{eq:finite-mass-charge-combinations}), these moments are
\begin{equation}
D_Q^i=2\eta m\Delta q\,r^i,\qquad
I_Q^{ij}=2\eta m\bar q\,r^{\langle ij\rangle},
\end{equation}
where angular brackets denote the symmetric trace-free (STF) projection. The $n$th scalar harmonic has frequency $n\Omega$. Its velocity factor and massive wave number are
\begin{equation}
\label{eq:massive-radiation-kinematics}
v_n=\sqrt{1-\frac{\mu^2}{n^2\Omega^2}},\qquad
k_n=n\Omega v_n.
\end{equation}
The radiative sector includes the leading contribution from each retained scalar multipole. It does not constitute a complete expansion of scalar corrections at a common relative PN order.
A scalar harmonic propagates only when $n\Omega>\mu$. We impose this source-side condition with the Heaviside step function $\Theta(n\Omega-\mu)$. Below threshold, the corresponding free mode is evanescent. Detector visibility is a separate question~\cite{Alsing2012MassiveBD,Diedrichs2024MassivePhase}.

At the same accuracy, the total flux is
\begin{equation}
\label{eq:leading-fluxes}
\begin{aligned}
\mathcal F={}&\mathcal F_T+\mathcal F_\phi^{\rm dip}+\mathcal F_\phi^{\rm quad},\\
\mathcal F_T={}&\frac{32}{5}\eta^2x^5
\left[1+\frac{4\kappa}{3}(1+y)\mathrm{e}^{-y}\right],\\
\mathcal F_\phi^{\rm dip}={}&\frac{1}{3}\eta^2(\Delta q)^2x^4v_1^3
\Theta(\Omega-\mu),\\
\mathcal F_\phi^{\rm quad}={}&\frac{16}{15}\eta^2\bar q^2x^5v_2^5
\Theta(2\Omega-\mu).
\end{aligned}
\end{equation}
The contribution $\mathcal F_T$ is the leading tensor quadrupole flux, computed with the GR quadrupole formula on the scalar-modified circular orbit~\cite{PetersMathews1963}. Its unmodified GR value is denoted by $\mathcal F_{\rm GR}=32\eta^2x^5/5$. The modified orbit produces a relative difference of a few parts in \mbox{$\displaystyle 10^4$} for the parameters shown. The tensor term does not include direct higher-PN modified-gravity tensor-generation effects. The terms $\mathcal F_\phi^{\rm dip}$ and $\mathcal F_\phi^{\rm quad}$ are the leading scalar dipole and scalar quadrupole fluxes. Their multipole normalization, angular integration, and massive phase-space factors are derived in \ref{app:worldline-radiation} and agree with the standard massive-scalar threshold structure~\cite{Alsing2012MassiveBD,Diedrichs2024MassivePhase,Shiralilou2022Waves,Liu2020MassiveBDWaveforms}. The quadrupole channel opens first at $2\Omega=\mu$, while the dipole channel opens later at $\Omega=\mu$. The dipole vanishes for $\Delta q=0$, whereas the scalar quadrupole generally remains.

At fixed $\beta_0$, the additional mass combinations in Fig.~\ref{fig:energy-flux}(a) illustrate the dependence on mass asymmetry. The $30\,\mathrm{M}_\odot+20\,\mathrm{M}_\odot$ system remains the benchmark for the charge values and local scalar-mass range quoted above. The $30\,\mathrm{M}_\odot+10\,\mathrm{M}_\odot$ curve probes a stronger local coupling and is included only as an illustrative comparison. The velocity factors suppress each scalar channel continuously near its opening threshold in Fig.~\ref{fig:energy-flux}(b). The tensor-flux comparison in Fig.~\ref{fig:energy-flux}(c) remains close to the GR value because the scalar-modified orbit changes the leading tensor quadrupole only perturbatively.

\subsection{Inspiral evolution and phase}
\label{sec:inspiral-phase}

The conservative energy and total flux from Sec.~\ref{sec:conservative-dynamics} determine the source-domain inspiral. In the adiabatic quasicircular approximation, their relation is the standard balance law~\cite{Shiralilou2022Waves}
\begin{equation}
\label{eq:energy-balance}
\frac{\mathrm{d}E}{\mathrm{d}t}=-\mathcal F.
\end{equation}
The PN variable $x$ provides a single evolution coordinate. Combining Eq.~(\ref{eq:energy-balance}) with Eqs.~(\ref{eq:binding-energy-frequency}) and~(\ref{eq:leading-fluxes}), and retaining only $\mathcal O(q_A^2)$ terms, gives
\begin{equation}
\label{eq:chirp-evolution}
\begin{aligned}
\dot x&=\frac{64}{5}\frac{\eta}{m}x^5\left[1+\delta_{\rm c}(x)+\delta_{\rm d}(x)+\delta_{\rm q}(x)\right]+\mathcal O(q_A^4),\\
\delta_{\rm c}&=\frac{2\kappa}{3}(1+y+2y^2)\mathrm{e}^{-y},\\
\delta_{\rm d}&=\frac{5}{96}(\Delta q)^2x^{-1}v_1^3\Theta(\Omega-\mu),\\
\delta_{\rm q}&=\frac{1}{6}\bar q^2v_2^5\Theta(2\Omega-\mu).
\end{aligned}
\end{equation}
The labels $\mathrm{c}$, $\mathrm{d}$, and $\mathrm{q}$ identify the finite-range conservative, scalar-dipole, and scalar-quadrupole contributions. The three terms enter the accumulated phase with distinct dependence on $x$ and $\mu$. The conservative contribution $\delta_{\rm c}$ combines the change in the tensor quadrupole flux evaluated on the scalar-modified circular orbit with the corresponding correction from the derivative of the binding energy.

At small $y$, the conservative kernel obeys $(1+y+2y^2)\mathrm{e}^{-y}=1+3y^2/2-5y^3/3+\mathcal O(y^4)$. The retained conservative contribution to the chirp therefore contains no linear range correction. Away from channel-opening thresholds, the massive velocity factors also begin quadratically in the scalar mass. With the leading matched charges, these finite-range and kinematic corrections produce no frequency-dependent linear-in-mass contribution to the source-domain phase.

Inverting the chirp relation in Eq.~(\ref{eq:chirp-evolution}) must preserve the same truncation. Define $\delta(x)\equiv\delta_{\rm c}(x)+\delta_{\rm d}(x)+\delta_{\rm q}(x)$. The linearized reciprocal used below additionally requires $|\delta(x)|\ll1$ over the frequency interval being interpreted quantitatively. The reciprocal is therefore expanded as $1/[1+\delta(x)]=1-\delta(x)+\mathcal O(q_A^4)$, which gives
\begin{equation}
\label{eq:time-orbital-phase-evolution}
\begin{aligned}
\frac{\mathrm{d}t}{\mathrm{d}x}&=\frac{5m}{64\eta}x^{-5}\left[1-\delta(x)\right]+\mathcal O(q_A^4),\\
\frac{\mathrm{d}\Phi}{\mathrm{d}x}&=\frac{5}{64\eta}x^{-7/2}\left[1-\delta(x)\right]+\mathcal O(q_A^4).
\end{aligned}
\end{equation}
The phase $\Phi$ is the orbital phase. The finite-mass Yukawa factor, velocity factors, and threshold gates remain intact in Eq.~(\ref{eq:time-orbital-phase-evolution}).

For the $n$th harmonic, let $f$ denote a positive Fourier frequency. We use $\tilde h(f)=\int h(t)\mathrm{e}^{-2\pi ift}\,\mathrm{d}t$. The stationary-phase approximation (SPA)~\cite{Droz1999SPA} selects $n\Omega_f=2\pi f$, or $\Omega_f=2\pi f/n$, and $x_f^{(n)}=(2\pi mf/n)^{2/3}$. The source Fourier phase is
\begin{equation}
\label{eq:source-fourier-phase}
\Psi_n(f)=n\Phi\!\left(x_f^{(n)}\right)-2\pi f\,t\!\left(x_f^{(n)}\right)+\frac{\pi}{4}.
\end{equation}
The source Fourier phase of the dominant tensor harmonic is $\Psi_2(f)$, with $x_f^{(2)}=(\pi mf)^{2/3}$. It contains neither a propagation phase nor a distance-dependent delay.

For finite $\mu$, we express the phase correction as a one-dimensional integral. The modified-minus-GR histories $\delta t$ and $\delta\Phi$ share the upper reference $x_{\rm ref}=0.1$, which is the endpoint of the working inspiral range. The numerical prescription imposes $\delta t(x_{\rm ref})=\delta\Phi(x_{\rm ref})=0$, which yields
\begin{equation}
\label{eq:finite-mass-phase-integral}
\scalebox{0.97}{$\displaystyle
\Delta\Psi_n(f)=\frac{5n}{64\eta}\int_{x_f^{(n)}}^{x_{\rm ref}}
\Bigl[x'^{-7/2}-\left(x_f^{(n)}\right)^{3/2}x'^{-5}\Bigr]\delta(x')\,\mathrm{d}x'
$}
\end{equation}
This one-dimensional phase integral retains the finite-range Yukawa factor, the velocity factors $v_1$ and $v_2$, and both threshold functions without a global PN power expansion. The common reference fixes the affine freedom in time and phase. It sets $\Delta\Psi_n=0$ and $\mathrm{d}\Delta\Psi_n/\mathrm{d}f=0$ when $x_f^{(n)}=x_{\rm ref}$. The numerical results follow from direct evaluation of this integral.

\begin{figure}[!ht]
\includegraphics[width=\columnwidth]{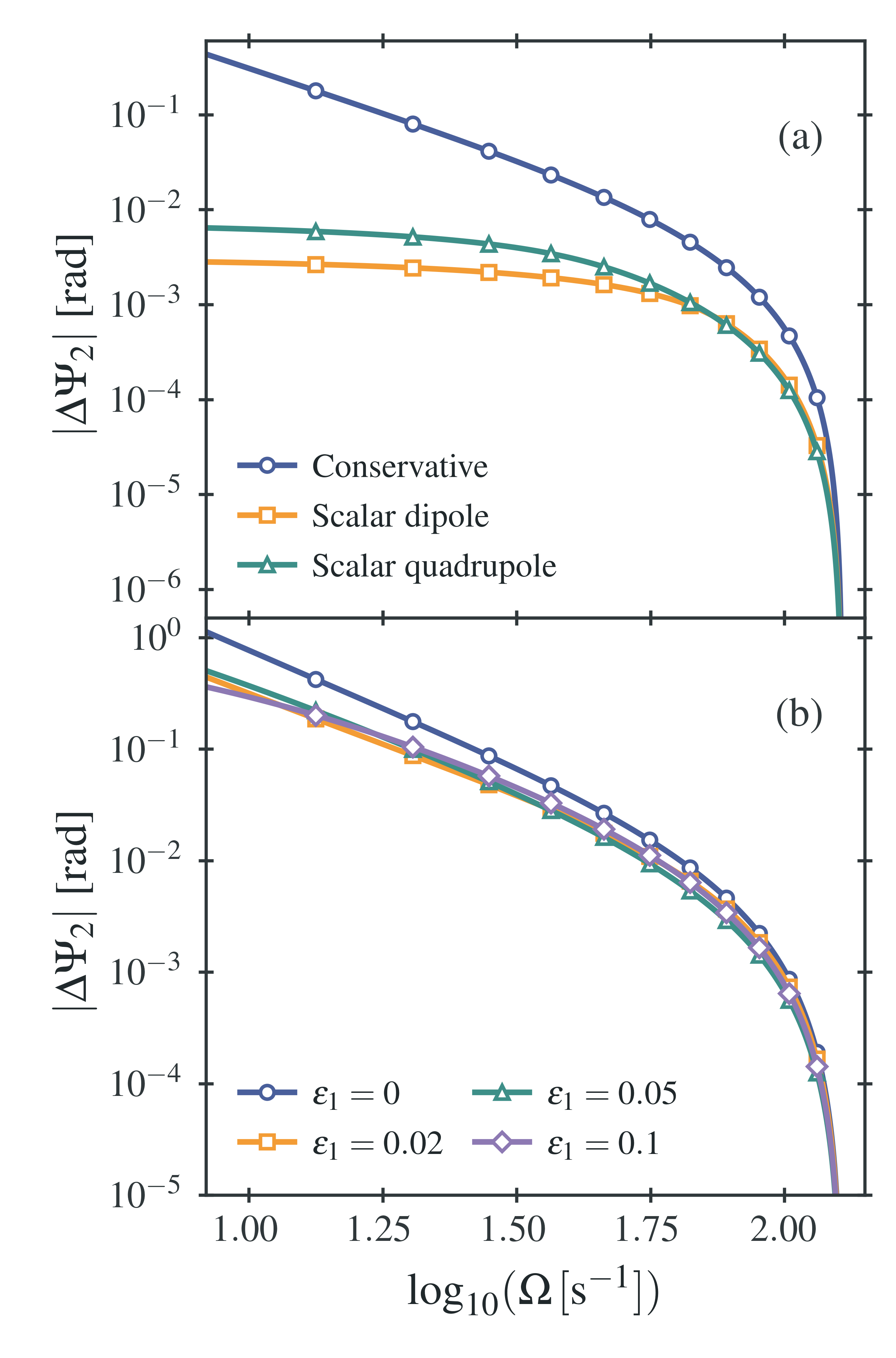}
\caption{Accumulated source-domain Fourier-phase corrections for the benchmark of Fig.~\ref{fig:conservative-dynamics}. (a) Magnitudes of the conservative, scalar-dipole, and scalar-quadrupole contributions to $\Delta\Psi_2$ at $\epsilon_1=0.020$. (b) Magnitude of the net correction for $\epsilon_1=0$, $0.02$, $0.05$, and $0.1$. The common reference is $x_{\rm ref}=0.1$, and the absolute value omits the phase sign.}
\label{fig:chirp-phase}
\end{figure}

In the massless limit, after omitting the affine constant and linear-in-frequency terms fixed by the chosen reference convention, the dipole contribution reduces to the following expression. As $\mu\to0$, $v_1,v_2\to1$ and $\mathrm{e}^{-y}\to1$,
\begin{equation}
\Delta\Psi_2^{\rm dip}(f)=\frac{5(\Delta q)^2}{7168\eta}(\pi mf)^{-7/3}.
\end{equation}
This source-phase result uses the convention of Eq.~(\ref{eq:source-fourier-phase}) and provides the standard $-1$PN dipole check. The conservative and scalar-quadrupole terms also contribute at 0PN order in this limit. Their complete finite-mass behavior remains encoded in Eq.~(\ref{eq:finite-mass-phase-integral})~\cite{Shiralilou2022Waves,Diedrichs2024MassivePhase}.

The dephasing in Fig.~\ref{fig:chirp-phase} accumulates over the inspiral history. For the benchmark shown in Fig.~\ref{fig:chirp-phase}(a), the conservative finite-range contribution has the largest magnitude throughout the displayed interval, while the relative scalar-dipole and scalar-quadrupole contributions vary with frequency. The massive cases shown have smaller net dephasing than the massless benchmark, but their ordering depends on frequency. The curve crossings reflect the competing effects of the finite-range conservative kernel and the radiation thresholds.

The phase integral retains the threshold functions~\cite{Alsing2012MassiveBD,Diedrichs2024MassivePhase}, although the sharp channel-opening prescription and ordinary SPA are nonuniform near $n\Omega=\mu$. The displayed continuation shows the limiting behavior of the retained approximation, with this narrow neighborhood excluded from quantitative waveform interpretation. A threshold-uniform treatment is left for future work.

\section{Scalar waveform and propagation}
\label{sec:scalar-waveform}

\subsection{Scalar waveform and tidal response}

\begin{figure*}[!t]
\begin{minipage}{\textwidth}
\centering
\includegraphics[width=0.95\textwidth]{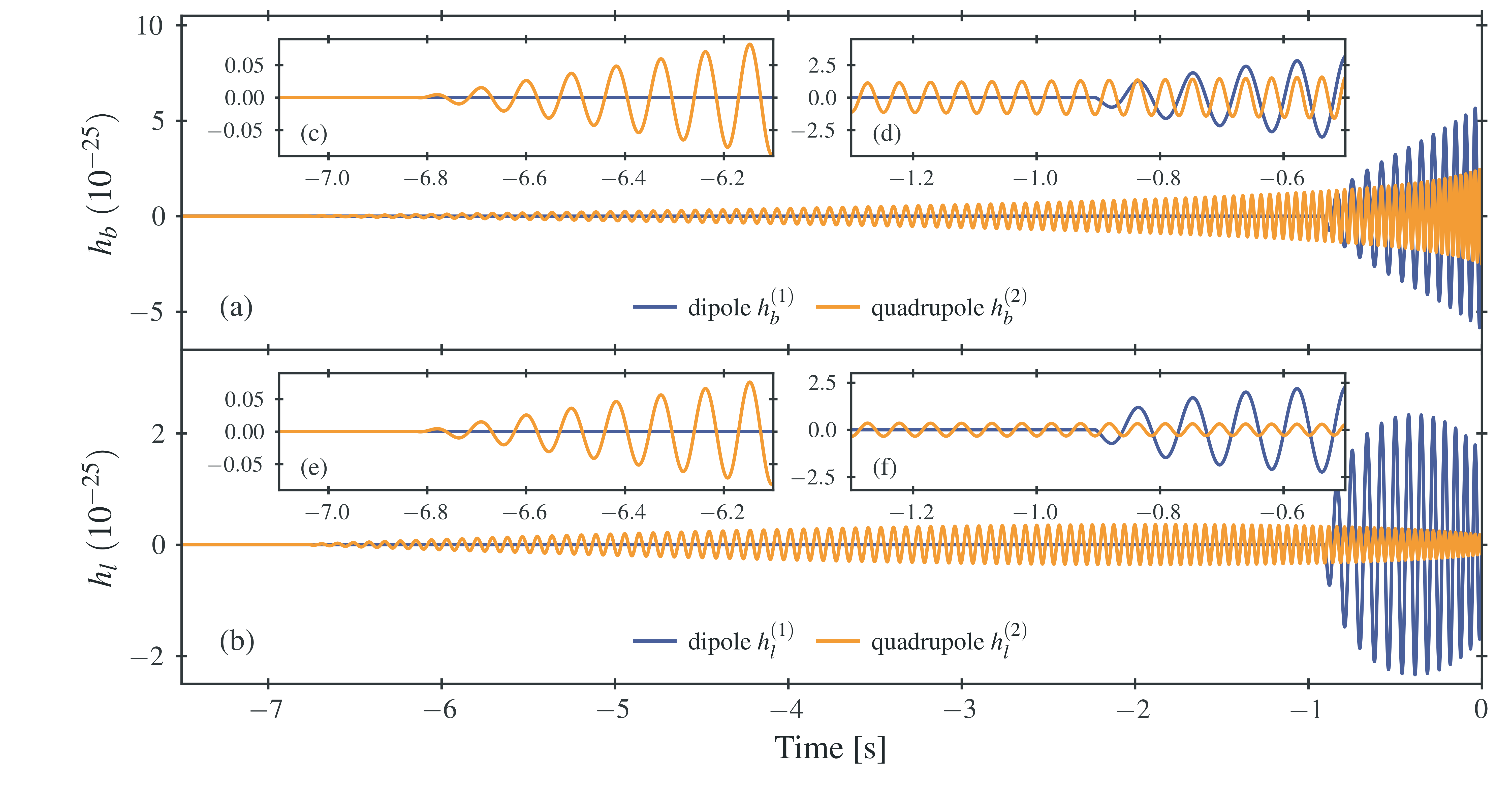}
\caption{Scalar waveforms for the binary benchmark and coupling of Fig.~\ref{fig:conservative-dynamics}, with $\epsilon_1=0.020$, $\alpha_0=0.1$, $\iota=\pi/3$, and $D=500\,\mathrm{Mpc}$. The $7.5\,\mathrm{s}$ interval ends at $x_{\rm ref}=0.1$. (a) The $n=1$ and $n=2$ contributions to $h_b$. (b) The corresponding contributions to $h_l$. Insets (c)--(f) magnify early and late portions of the same signals. The individual harmonic contributions are displayed before dispersive propagation. The distance $D$ enters only through the geometric $1/D$ amplitude normalization.}
\label{fig:scalar-waveforms}
\end{minipage}
\end{figure*}

Sec.~\ref{sec:binary-dynamics} determines the orbital history and Fourier phase, while \ref{app:worldline-radiation} fixes the radiative scalar-field normalization from the matched charges. Let $D$ denote the source--observer distance within the asymptotically flat propagation model. For a circular orbit, take $\mathbf r=r(\cos\Phi,\sin\Phi,0)$ and $\mathbf N=(\sin\iota,0,\cos\iota)$, where $\cos\iota=\mathbf N\cdot\hat{\mathbf L}$ and $\hat{\mathbf L}$ is the unit orbital-angular-momentum vector. Substitution of the source multipoles into the radiative field gives the two propagating scalar harmonics
\begin{equation}
\label{eq:scalar-time-domain-harmonics}
\begin{aligned}
\phi_1(t;D)&=-\frac{2\eta m\Delta q}{D}x^{1/2}v_1\sin\iota\sin\Phi(t)\Theta(\Omega-\mu),\\
\phi_2(t;D)&=-\frac{2\eta m\bar q}{D}xv_2^2\sin^2\iota\cos\!\left[2\Phi(t)\right]\Theta(2\Omega-\mu).
\end{aligned}
\end{equation}
The $n=1$ field is controlled by $\Delta q$, whereas the $n=2$ field is controlled by $\bar q$. The factors $v_1$ and $v_2^2$ drive the radiative amplitudes to zero at their respective thresholds, while $1/D$ gives the geometric radiative decay. The threshold functions select the propagating radiative channels. Massive-scalar binary waveform calculations motivate the separation between the radiative harmonics and the subsequent propagation problem~\cite{Liu2020MassiveBDWaveforms,Higashino2023HorndeskiWaveforms}.

For the $n$th harmonic, write $\omega_n=n\Omega$ and $k_n=\sqrt{\omega_n^2-\mu^2}$. The real time-domain fields use the convention $\phi=\Re[\phi_{\omega_n}\mathrm{e}^{i(\omega_n t-k_n\mathbf N\cdot\mathbf x)}]$. The complex representation used in \ref{app:worldline-radiation} is the conjugate of the positive-frequency representation adopted here, so the two yield the same real field. The charges $Q_A$ and $q_A$ control scalar generation, whereas $\alpha_0$ converts the canonical scalar field into a metric response.

Around the local Minkowski background, the scalar part of the physical metric perturbation is $h^{(s)}_{\mu\nu}=\alpha_0\phi\eta_{\mu\nu}$. The physical scalar strain in this relation follows from the adopted leading conformal coupling $\alpha_0$~\cite{Evstafyeva2023ScalarPolarization}. Let $P_{ij}\equiv\delta_{ij}-N_iN_j$ denote the transverse projector. The linearized tidal tensor and the tidal-equivalent polarizations obey
\begin{equation}
\label{eq:scalar-tidal-response}
\begin{aligned}
R^{(s)}_{0i0j}&=\frac{\alpha_0}{2}\left(\omega_n^2P_{ij}+\mu^2N_iN_j\right)\phi_n,\\
R^{(s)}_{0i0j}&=-\frac{1}{2}\partial_t^2\left(h_b^{(n)}P_{ij}+h_l^{(n)}N_iN_j\right),\\
h_b^{(n)}&=\alpha_0\phi_n,\\
h_l^{(n)}&=\frac{\mu^2}{n^2\Omega^2}h_b^{(n)}=(1-v_n^2)h_b^{(n)}.
\end{aligned}
\end{equation}
The identity $1-v_n^2=\mu^2/(n^2\Omega^2)$ fixes the relative longitudinal response and identifies $h_b^{(n)}$ and $h_l^{(n)}$ as two tidal components generated by the same massive scalar degree of freedom. Close to the opening threshold of a given harmonic, $h_l^{(n)}/h_b^{(n)}$ approaches unity, while the radiative scalar amplitude vanishes through the relevant $v_n$ or $k_n$ factor. This tidal interpretation agrees with the standard massive-scalar polarization analysis~\cite{Hou2018Polarizations}. An instrument-specific detector response requires the chosen time-delay-interferometry combination and its sensitivity to nontensorial polarizations~\cite{my_paper_TDI}.

The waveforms show the expected chirping behavior, with increasing oscillation frequencies and growing breathing-mode amplitudes toward the adopted inspiral endpoint. By contrast, the longitudinal envelopes need not grow monotonically because $\mu^2/(n^2\Omega^2)$ decreases during the inspiral. The two harmonics remain distinguishable through their phase evolution and frequency content, while their longitudinal components follow the same timing as the breathing components with the rescaling in Eq.~(\ref{eq:scalar-tidal-response}). Superposition of the dipole and quadrupole pieces can therefore produce visible interference.

\begin{figure}[!ht]
\centering
\includegraphics[width=0.96\columnwidth]{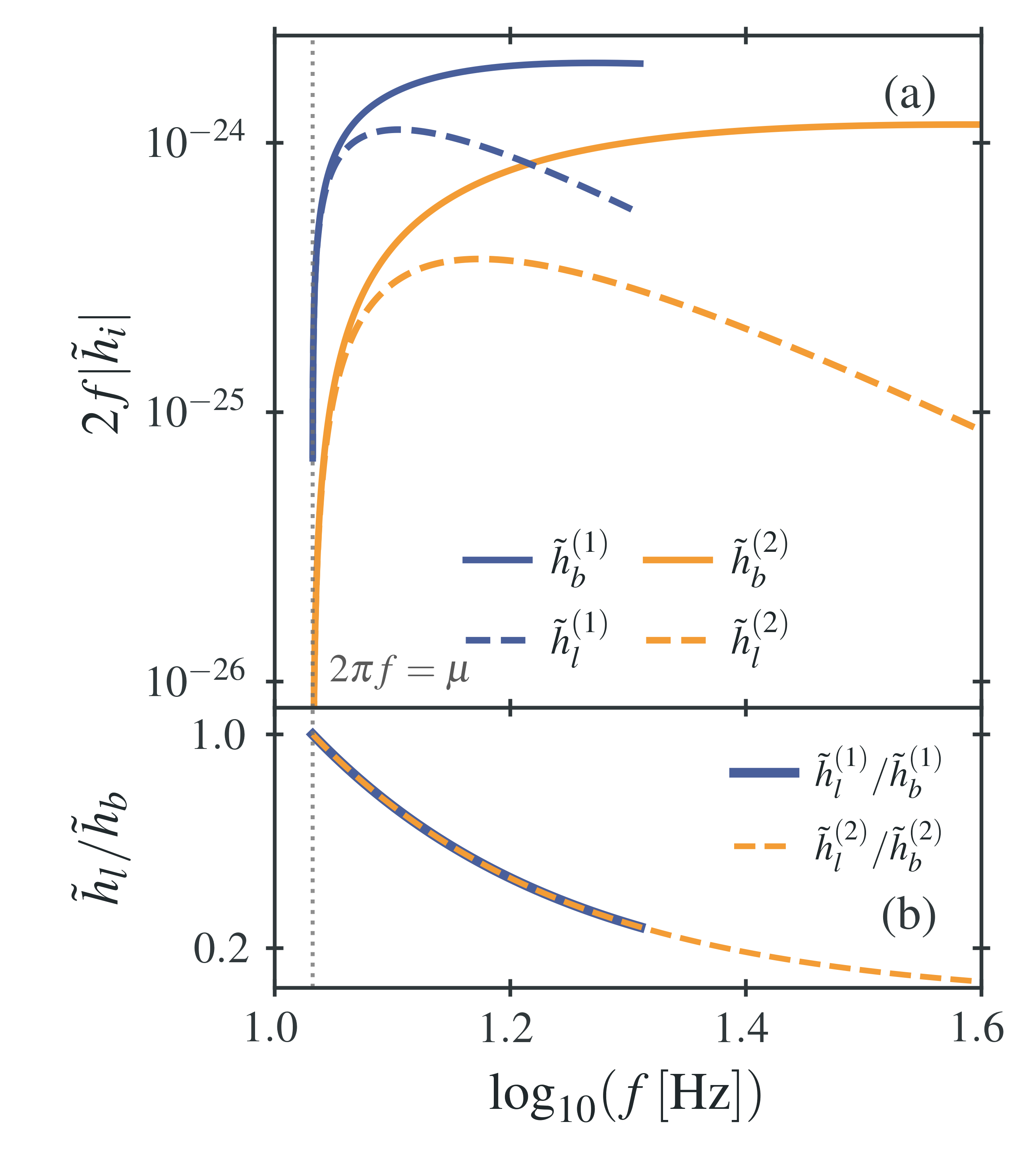}
\caption{Frequency-domain scalar waveform for the source and geometric parameters of Fig.~\ref{fig:scalar-waveforms}, before distance-dependent dispersive propagation. Blue and orange denote the $n=1$ and $n=2$ harmonics, respectively. (a) Frequency-weighted amplitudes \mbox{$\displaystyle 2f|\tilde h_i|$}, with \mbox{$\displaystyle i=b,l$}. Solid and dashed curves show the breathing and longitudinal responses, respectively. (b) Longitudinal-to-breathing amplitude ratios for both harmonics. The vertical gray dotted line marks the propagation threshold $2\pi f=\mu$.}
\label{fig:frequency-domain-scalar-waveform}
\end{figure}

For the frequency-domain scalar waveform, the ordinary SPA is applied directly to Eq.~(\ref{eq:scalar-time-domain-harmonics}). Define the leading GR orbital chirp rate by $\dot\Omega_{\rm GR}(x)=96\eta x^{11/2}/(5m^2)$ and the common Fourier-frequency velocity factor by $v_f\equiv\sqrt{1-\mu^2/(2\pi f)^2}$. The resulting restricted-amplitude expressions are
\begin{equation}
\begin{aligned}
\tilde h_b^{(1)}(f)&=+i\frac{\alpha_0\eta m\Delta q}{D}\frac{[x_f^{(1)}]^{1/2}v_f\sin\iota}{\sqrt{\dot\Omega_{\rm GR}(x_f^{(1)})/(2\pi)}}\\
&\quad \times\Theta(2\pi f-\mu)\mathrm{e}^{i\Psi_1(f)},\\
\tilde h_b^{(2)}(f)&=-\frac{\alpha_0\eta m\bar q}{D}\frac{x_f^{(2)}v_f^2\sin^2\iota}{\sqrt{\dot\Omega_{\rm GR}(x_f^{(2)})/\pi}}\\
&\quad \times\Theta(2\pi f-\mu)\mathrm{e}^{i\Psi_2(f)},\\
\tilde h_l^{(n)}(f)&=(1-v_f^2)\tilde h_b^{(n)}(f).
\end{aligned}
\end{equation}
The stationary mapping evaluates the first two harmonics at $n\Omega_f=2\pi f$. Their distinct time-domain thresholds therefore become the common Fourier gate $\Theta(2\pi f-\mu)$. The source phase retains the corrections of Sec.~\ref{sec:inspiral-phase}, while the GR chirp rate is sufficient in the scalar amplitude Jacobian at the retained order~\cite{Liu2020MassiveBDWaveforms,Higashino2023HorndeskiWaveforms}. At fixed \mbox{$\displaystyle f$}, the two harmonic ratios in Fig.~\ref{fig:frequency-domain-scalar-waveform}(b) coincide because the same massive-scalar kinematics determines the longitudinal-to-breathing conversion.

The two harmonics share the same orbital endpoint, but their Fourier-frequency endpoints differ because the stationary mapping uses $n\Omega_f=2\pi f$. Their source amplitudes also carry different powers of $x_f^{(n)}$ and different charge combinations. The frequency-domain result therefore combines a common massive-scalar propagation gate with harmonic-dependent source support and amplitude scaling.

We use the leading GR amplitudes for the tensor plus and cross polarizations $h_+$ and $h_\times$, with the orbital history and $\Psi_2(f)$ evaluated in the same source-phase convention. The orbit-induced amplitude correction is omitted in this restricted-amplitude description.

\subsection{Massive propagation}

The radiation variables of Sec.~\ref{sec:conservative-dynamics} already contain the massive kinematics. The velocity factor $v_n$ in Eq.~(\ref{eq:massive-radiation-kinematics}) is the group velocity of the propagating scalar harmonic. With $\omega_n=n\Omega$ and $k_n$ defined above, the local dispersion relation is
\begin{equation}
\omega_n^2=k_n^2+\mu^2.
\end{equation}
The two harmonics therefore open at different orbital frequencies. The quadrupole branch opens first at $2\Omega=\mu$, whereas the dipole branch opens at $\Omega=\mu$.

The propagation model assumes an asymptotically flat spacetime, with cosmological expansion and redshift left for future work. For the outgoing $n$th-harmonic mode $\exp[i(\omega_n t-k_nD)]$, the propagation transfer retains the full dependence on $\mu/\omega_n$. Relative to massless propagation over the same distance, with the common light-travel time removed, the excess frequency-domain phase is
\begin{equation}
\label{eq:propagation-phase}
\Delta\Psi_{\rm prop}(f)=D\left[2\pi f-\sqrt{(2\pi f)^2-\mu^2}\right].
\end{equation}
For $2\pi f>\mu$, this expression is equivalent to $D(\omega_n-k_n)$ after the stationary mapping $\omega_n=n\Omega=2\pi f$. It gives each open scalar harmonic the multiplicative factor $\mathrm{e}^{i\Delta\Psi_{\rm prop}(f)}$,
\begin{equation}
\tilde h_{b,l,{\rm prop}}^{(n)}(f)=\tilde h_{b,l}^{(n)}(f)\mathrm{e}^{i\Delta\Psi_{\rm prop}(f)}.
\end{equation}
The derivative check $-\partial_\omega\Delta\Psi_{\rm prop}=D(1/v_f-1)>0$ gives the corresponding excess propagation time. Far above threshold, $2\pi f\gg\mu$ gives $\Delta\Psi_{\rm prop}(f)\simeq D\mu^2/(4\pi f)$, so the dispersive correction decreases as the wave frequency rises. Below $2\pi f=\mu$, the free massive mode is evanescent and is excluded from the propagating waveform~\cite{Alsing2012MassiveBD,Diedrichs2024MassivePhase}.

\begin{figure}[!ht]
\includegraphics[width=\columnwidth]{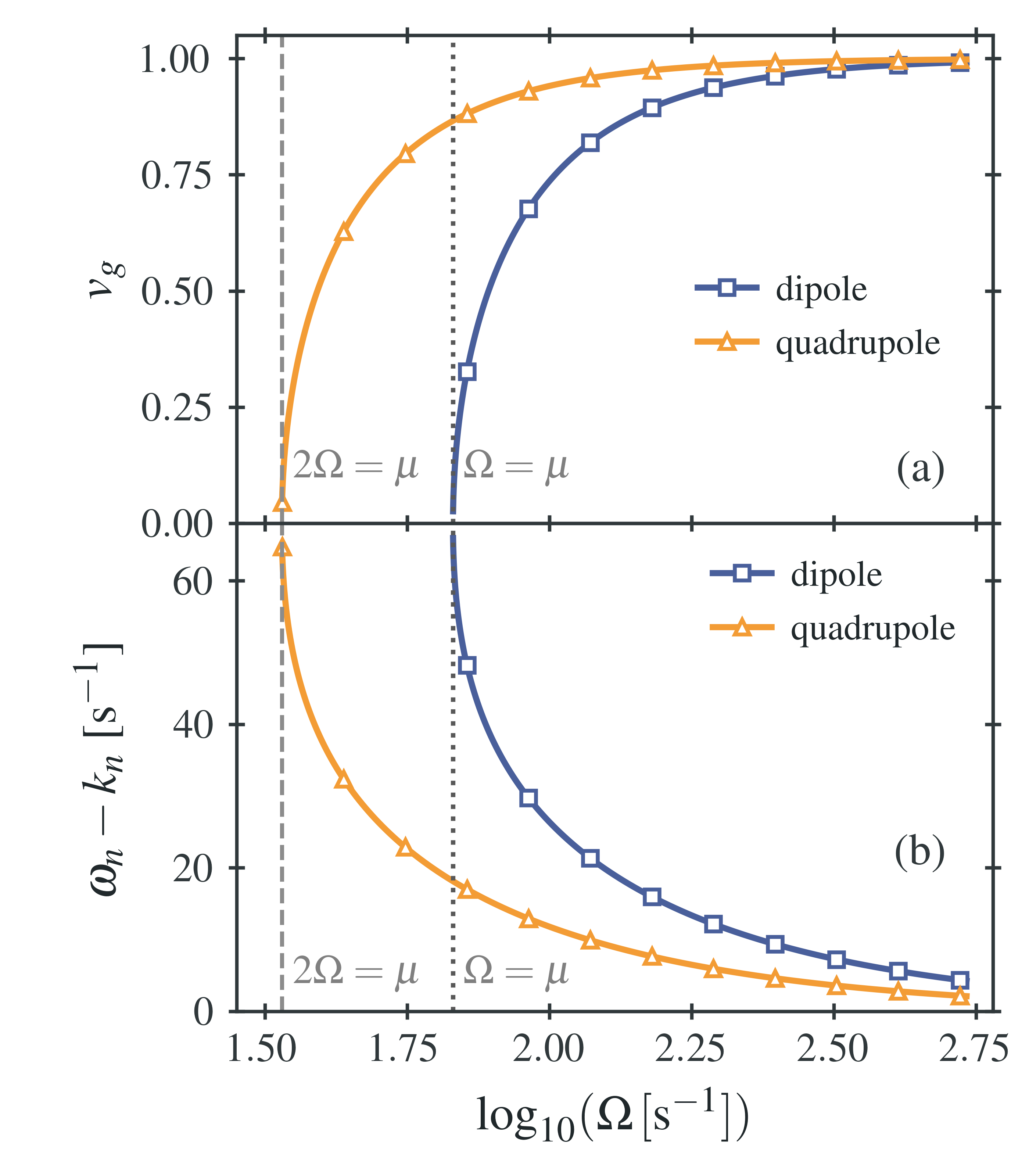}
\caption{Local massive-scalar dispersion relations for the benchmark and scalar mass of Fig.~\ref{fig:scalar-waveforms}. (a) Group velocities of the dipole and quadrupole branches. (b) \mbox{$\displaystyle \omega_n-k_n$} for the dipole and quadrupole harmonics. The vertical gray dashed and dotted lines identify the $2\Omega=\mu$ and $\Omega=\mu$ thresholds, respectively, which separate evanescent and propagating regimes.}
\label{fig:massive-propagation}
\end{figure}

The group velocity in Fig.~\ref{fig:massive-propagation}(a) approaches zero at the opening threshold and tends to unity far above threshold. The $n=2$ branch opens at lower orbital frequency because its scalar harmonic has twice the orbital frequency. At fixed observed Fourier frequency $f$, \mbox{$\displaystyle \omega_n-k_n$} is determined by $f$ and $\mu$, while the harmonic distinction enters through the source mapping and the orbital-frequency thresholds. Multiplication of \mbox{$\displaystyle \omega_n-k_n$} by \mbox{$\displaystyle D$} gives the propagation phase in Eq.~(\ref{eq:propagation-phase}).

\section{Discussion and conclusions}
\label{sec:discussion}

We connect the curvature-sourced scalar response of an isolated BH to BBH inspiral signals in a massive sGB effective theory. The static calculation determines the second-order local coefficient and its finite conversion to a renormalized worldline coupling. The binary application uses the leading matched charges in the finite-range dynamics, scalar radiation, source-domain phase, and breathing and longitudinal responses. We then apply massive propagation as a separate kinematic effect.

The local static coefficient is obtained from the complete curvature-sourced response through overlap matching. The linear finite-mass correction cancels, and the first nonvanishing contribution to the matched local coefficient is the analytic quadratic term $-4\epsilon^2/3$. An independent numerical reconstruction of the perturbative radial problem is consistent with this coefficient. Matching to the point-source field includes a finite curvature contribution, whose scale dependence cancels that of the renormalized worldline coefficient in the matched monopole sector. Away from thresholds, the finite-mass correction to the source-domain phase also begins quadratically at the retained order. The scalar mass also introduces a finite Yukawa range and distinct dipole and quadrupole radiation thresholds. These effects modify the orbital evolution and accumulate in the source-domain phase.

The results apply within a controlled weak-coupling treatment of nonspinning BHs on quasicircular inspirals, with a small local scalar-mass parameter and leading-PN binary dynamics. Within this domain, the calculation retains the finite Yukawa range, massive radiation thresholds, the leading matched charges, and asymptotically flat dispersive propagation.

Future work can extend the charge matching to higher order and develop higher-PN dynamics for spinning, eccentric, and precessing binaries. The second-order static coupling can be combined with the corresponding weak-zone interactions and radiation in a common higher-PN calculation. Perturbations from the disk of an active galactic nucleus (AGN) and hierarchical motion around a Kerr supermassive BH can modulate the phase of a stellar-mass BBH~\cite{my_paper_AGN,my_paper_Kerr}. These environmental and hierarchical effects must be distinguished from the finite-mass scalar effects considered here. Further work is also needed for threshold-uniform waveforms, merger and ringdown modeling, cosmological propagation, and detector-level inference.

\section*{Acknowledgements}
This work was supported by the National Key Research and Development Program of China (Grant No. 2023YFC2206702), the National Natural Science Foundation of China (Grants No. 125B2102, No. 12575072, and No. 12547101), and the Fundamental Research Funds for the Central Universities Project (Grant No. 2024IAIS-ZD009).

\section*{Data Availability}
No external data were used or analyzed in this study. The numerical results and figures were generated directly from the equations presented in the paper.

\appendix

\section{Technical details of the scalar matching}
\label{app:scalar-matching}

We use the isolated Schwarzschild mass $M$ throughout this appendix, with $r_h=2M$ and $\epsilon=2\mu M$ also after the dimensionless reduction. The dimensionless radial equation in Eq.~(\ref{eq:dimensionless-radial-equation}) follows from the scaled Schwarzschild radial problem. Expanding it around $z=1$ and selecting the regular Frobenius branch gives the leading horizon relation
\begin{equation}
\label{eq:horizon-regularity-relation}
\psi'(1)-\epsilon^2\psi(1)=-12.
\end{equation}
This relation fixes the regular horizon derivative once the horizon value is selected.

The massless solution and the first regular near-zone correction provide the local side of the overlap. At fixed $z$, we separate the locally determined particular sector from the horizon-regular homogeneous constant mode. The locally determined part is organized as $\psi_0+\epsilon^2\psi_2^{\rm reg}+\cdots$, while the coefficient of the regular homogeneous constant is fixed only by matching and may carry nonanalytic dependence on $\epsilon$. The $\mathcal O(\epsilon^2)$ near equation is
\begin{equation}
\label{eq:appendix-near-equation}
\frac{\mathrm{d}}{\mathrm{d}z}\left[z(z-1)\frac{\mathrm{d}\psi_2}{\mathrm{d}z}\right]=z^2\psi_0.
\end{equation}
Integrating once and imposing horizon regularity gives
\begin{equation}
z(z-1)(\psi_2^{\rm reg})'=2z^2+2z-4+\frac43\log z.
\end{equation}
With $\psi_2^{\rm reg}(1)=0$, a horizon-regular particular solution is
\begin{equation}
\label{eq:appendix-near-solution}
\psi_2^{\rm reg}=2z+4\log z+\frac{4}{3}\operatorname{Li}_2\!\left(1-\frac{1}{z}\right)-2,
\end{equation}
where $\operatorname{Li}_2(x)$ denotes the dilogarithm. The horizon-regular homogeneous solution, normalized to unity at the horizon, has the expansion
\begin{equation}
1+\epsilon^2\left(\frac{z^2}{6}+\frac z3+\frac{\log z}{3}-\frac12\right)+\cdots.
\end{equation}
Its coefficient is fixed by matching to the decaying far solution. The terms needed for matching at large $z$ are
\begin{equation}
\begin{aligned}
\psi_2^{\rm reg}={}&2z+4\log z+\frac{2\pi^2}{9}-2
-\frac{4}{3}\frac{\log z+1}{z}\\
&-\frac{2}{3}\frac{\log z}{z^2}-\frac{1}{3z^2}+\cdots.
\end{aligned}
\end{equation}

After separating the logarithmic structure from the pure monopole coefficient, the locally determined $z^{-1}$ sector is
\begin{equation}
\left.\psi\right|_{z^{-1}\ {\rm sector}}
=
\frac{4-\frac{4}{3}\epsilon^2}{z}
-
\frac{4\epsilon^2}{3}\frac{\log z}{z}
+\cdots .
\end{equation}
The first term isolates the pure $1/z$ coefficient entering the stated matching prescription, while the logarithmic term remains part of the overlap structure.

When an exponential times a power of \mbox{$\displaystyle z$} is substituted into the homogeneous radial equation, the terms proportional to \mbox{$\displaystyle z^2$} fix the exponential rates to \mbox{$\displaystyle \pm\epsilon$}. The terms proportional to \mbox{$\displaystyle z$} then fix the powers to \mbox{$\displaystyle -1-\epsilon/2$} for the decaying mode and \mbox{$\displaystyle -1+\epsilon/2$} for the growing mode.

The near and far descriptions are matched with $\rho=\epsilon z$. The scaling $\psi(z)=\epsilon\chi(\rho;\epsilon)$ organizes the Yukawa solution at $\rho=\mathcal O(1)$ while preserving the appropriate leading-amplitude normalization. The superscript $N$ denotes the far solution normalized to unit decaying amplitude at infinity. Substituting these variables into Eq.~(\ref{eq:dimensionless-radial-equation}) gives
\begin{equation}
\frac{\mathrm{d}}{\mathrm{d}\rho}\left[(\rho^2-\epsilon\rho)\frac{\mathrm{d}\chi}{\mathrm{d}\rho}\right]-\rho^2\chi=-\frac{12\epsilon^3}{\rho^4}.
\end{equation}
Writing $\bar L_0[f]\equiv(\rho^2f')'-\rho^2f$ and $\bar L_1[f]\equiv(\rho f')'$, with primes denoting derivatives with respect to $\rho$, the normalized homogeneous far hierarchy is
\begin{equation}
\label{eq:far-zone-hierarchy}
\begin{aligned}
\bar L_0[\chi_0]&=0,\\
\bar L_0[\chi_1^N]&=\bar L_1[\chi_0],\\
\bar L_0[\chi_2^N]&=\bar L_1[\chi_1^N],\\
\bar L_0[\chi_3^{\rm hom}]&=\bar L_1[\chi_2^N].
\end{aligned}
\end{equation}
The forced contribution first enters at order $\epsilon^3$ through
\begin{equation}
\bar L_0[\chi_3^{\rm p}]
=
-\frac{12}{\rho^4}.
\end{equation}
\begin{equation}
\begin{aligned}
\chi(\rho;\epsilon)={}&A_Y(\epsilon)
\biggl[
\chi_0(\rho)
+\epsilon\chi_1^N(\rho)\\
&\quad+\epsilon^2\chi_2^N(\rho)
+\epsilon^3\chi_3^{\rm hom}(\rho)
+\cdots
\biggr]\\
&+\epsilon^3\chi_3^{\rm p}(\rho)+\cdots .
\end{aligned}
\end{equation}
The normalized decaying hierarchy carries unit homogeneous amplitude and is multiplied by the physical far-zone amplitude $A_Y$. The curvature-forced particular layer is additive and is not multiplied by $A_Y$. The term proportional to $\epsilon\rho$ is the first Schwarzschild curvature correction and is retained in the overlap analysis. The leading decaying solution is $\chi_0(\rho)=\mathrm{e}^{-\rho}/\rho$. Factoring out $\chi_0$ and integrating the first two inhomogeneous equations gives
\begin{equation}
\begin{aligned}
\frac{\mathrm{d}}{\mathrm{d}\rho}\left(\frac{\chi_1^N}{\chi_0}\right)
&=-\mathrm{e}^{2\rho}\int_\rho^\infty\frac{\mathrm{e}^{-t}}{t}\bar L_1[\chi_0](t)\,\mathrm{d}t,\\
\frac{\mathrm{d}}{\mathrm{d}\rho}\left(\frac{\chi_2^N}{\chi_0}\right)
&=-\mathrm{e}^{2\rho}\int_\rho^\infty\frac{\mathrm{e}^{-t}}{t}\bar L_1[\chi_1^N](t)\,\mathrm{d}t.
\end{aligned}
\end{equation}
The operator inside each integral acts on the integration variable, and the upper limit excludes the growing mode. Expanding the unit decaying prefactor $\rho^{-\epsilon/2}$ fixes the remaining constants through
\begin{equation}
\begin{aligned}
\frac{\chi_1^N}{\chi_0}+\frac12\log\rho&\longrightarrow0,\\
\frac{\chi_2^N}{\chi_0}-\frac18\log^2\rho&\longrightarrow0,
\qquad \rho\longrightarrow\infty.
\end{aligned}
\end{equation}
The first integral yields $\chi_1^N(\rho)=\mathrm{e}^{-\rho}[1/(2\rho)-\mathrm{e}^{2\rho}E_1(2\rho)/2-\log\rho/2]/\rho$.
The function $E_1(x)\equiv\int_x^\infty \mathrm{e}^{-t}\mathrm{d}t/t$ is the exponential integral for $x>0$. The normalized second layer has the small-$\rho$ expansion
\begin{equation}
\begin{aligned}
\chi_2^N(\rho)={}&\frac{1}{3\rho^3}
+\frac{\gamma_{\rm E}+\log 2-1}{4\rho^2}\\
&+\frac{-\frac13\log\rho+\Lambda_2^N}{\rho}
+\mathcal O(\log\rho).
\end{aligned}
\end{equation}
The finite-part normalization constant $\Lambda_2^N$ is fixed by these integrals and the unit-amplitude boundary conditions. It is an intermediate far-zone normalization constant rather than a physical parameter and cancels from $C_{\rm loc}$.

The large-$z$ prefactor expansion and the fixed-$\rho$ hierarchy are two asymptotic orderings of the same normalized decaying solution. This hierarchy resums the prefactor terms that are nonuniform in the overlap region. Reexpressing the decaying basis function in Eq.~(\ref{eq:distorted-yukawa-basis}) in terms of $\rho$ yields
\begin{equation}
u_{\rm dec}(z)\sim\epsilon^{1+\epsilon/2}\mathrm{e}^{-\rho}\rho^{-1-\epsilon/2}.
\end{equation}
Comparing $\psi_{\rm hom}=C_{\rm dec}u_{\rm dec}$ with the normalized far form $\psi_{\rm hom}\sim\epsilon A_Y\mathrm e^{-\rho}\rho^{-1-\epsilon/2}$ gives $A_Y=C_{\rm dec}\epsilon^{\epsilon/2}$. The curvature conversion then removes the remaining linear contribution in the local coefficient.

The source-subtracted homogeneous and forced third layers must be kept distinct. The physical layer is the sum of the homogeneous contribution weighted by $A_Y$ and the forced particular contribution, with the normalized hierarchy defined by $\bar L_0$ and $\bar L_1$ above. Retaining only the $\rho^{-2}$ sector needed below, the third-layer coefficients are
\begin{equation}
\label{eq:far-third-layer-expansion}
\begin{aligned}
\left.\chi_3^{\rm hom}\right|_{\rho^{-2}}
&=\left[-\frac{1}{6}\log\rho+\frac{5}{24}+\frac{\Lambda_2^N}{2}\right]\frac{1}{\rho^2},\\
\left.\chi_3^{\rm p}\right|_{\rho^{-2}}
&=-\frac{1}{2\rho^2}.
\end{aligned}
\end{equation}
The term $-1/\rho^4$ in the particular solution belongs to the forced sector identified above and does not enter this $\rho^{-2}$ combination. At $A_Y(0)=4$, the $1/\rho^2$ coefficient in the unnormalized physical third-layer combination is $4(5/24+\Lambda_2^N/2)-1/2=1/3+2\Lambda_2^N$. Dividing by the leading homogeneous amplitude $4$ gives the normalized coefficient $1/12+\Lambda_2^N/2$. This coefficient enters the subsequent order-by-order matching. Since $\epsilon^4\rho^{-2}=\epsilon^2/z^2$ while $\epsilon^4\rho^{-1}=\epsilon^3/z$, the $\chi_3$ layer completes the nonmonopole $\mathcal O(\epsilon^2/z^2)$ consistency check but does not alter the $\mathcal O(\epsilon^2)/z$ local monopole. Including the corresponding lower and third-layer terms gives
\begin{equation}
-\frac{2}{3}\frac{\epsilon^2\log z}{z^2}-\frac{1}{3}\frac{\epsilon^2}{z^2},
\end{equation}
with all $\log\epsilon$ terms canceled. This agrees with the near expansion above.

After writing $\rho=\epsilon z$, the homogeneous contribution to the overlap monopole sector is
\begin{equation}
\begin{aligned}
\frac{A_Y}{z}\biggl[1&+\frac{\gamma_{\rm E}+\log 2-1}{2}\epsilon\\
&+\epsilon^2\left(-\frac13\log\epsilon-\frac13\log z+\Lambda_2^N\right)\biggr]+\cdots.
\end{aligned}
\end{equation}
Matching the $1/z$ and $(\log z)/z$ terms of the near and far expansions gives
\begin{equation}
\begin{aligned}
C_{\rm dec}={}&4+\epsilon\left[-2\log\epsilon+2-2\gamma_{\rm E}-2\log 2\right]\\
&+\epsilon^2\biggl[\frac12\log^2\epsilon
+\left(\frac13+\gamma_{\rm E}+\log 2\right)\log\epsilon\\
&-\frac43+(\gamma_{\rm E}+\log 2-1)^2-4\Lambda_2^N\biggr]\\
&+\mathcal O\!\left(\epsilon^3\log^p\epsilon\right).
\end{aligned}
\end{equation}
The basis conversion $A_Y=C_{\rm dec}\epsilon^{\epsilon/2}$ gives
\begin{equation}
\begin{aligned}
A_Y={}&4+\left(2-2\gamma_{\rm E}-2\log 2\right)\epsilon\\
&+\epsilon^2\biggl[\frac43\log\epsilon-\frac43\\
&+\left(\gamma_{\rm E}+\log 2-1\right)^2-4\Lambda_2^N\biggr]\\
&+\mathcal O\!\left(\epsilon^3\log^p\epsilon\right).
\end{aligned}
\end{equation}
The conversion to the local coefficient is
\begin{equation}
\begin{aligned}
C_{\rm loc}={}&A_Y\biggl[1+\frac{\gamma_{\rm E}+\log 2-1}{2}\epsilon\\
&+\epsilon^2\left(-\frac13\log\epsilon+\Lambda_2^N\right)\biggr]\\
&+\mathcal O\!\left(\epsilon^3\log^p\epsilon\right).
\end{aligned}
\end{equation}
All logarithmic pieces and $\Lambda_2^N$ cancel in this expression, reproducing the local-coefficient result in Eq.~(\ref{eq:local-coefficient-massless-limit}).

Let $z_m$ denote a matching point satisfying $1\ll z_m\ll\epsilon^{-1}$. With consistent coordinates and logarithmic terms, the complete $\mathcal O(\epsilon^2)$ near/far conversion cancels the $z_m$ dependence, leaving $C_{\rm loc}$ independent of the matching point through this order.

To match this static coefficient to a worldline source, we compare the BH and unit-source fields in the same isotropic radius $R$, where $r=R(1+M/2R)^2$. This matching separates the curvature correction to the point-source field from the local coupling~\cite{Julie2019Dynamics,Ivanov2023Matching}. We use dimensional regularization for the weak-field point-source calculation, with spatial dimension $d$ and $G_d=\ell^{d-3}$. The dimensional continuation serves only as a regulator. The isotropic Schwarzschild--Tangherlini exterior provides the weak-field expansion~\cite{Mougiakakos2021Schwarzschild}. Defining $\Omega_{d-1}=2\pi^{d/2}/\Gamma(d/2)$, $A_d=4\pi G_dM/[(d-1)\Omega_{d-1}]$, and \mbox{$\displaystyle a=A_d/R^{d-2}$}, the static scalar field per unit source obeys
\begin{equation}
\label{eq:unit-source-dr-operator}
\mbox{$\displaystyle \boldsymbol\nabla\!\cdot[(1-a^2)\boldsymbol\nabla U]-\mu^2(1-a)(1+a)^{1+4/(d-2)}U=0.$}
\end{equation}
The operator is expanded in the weak field before regularization. With \mbox{$\displaystyle b_d=4\pi G_d/[(d-2)\Omega_{d-1}]$}, the unit-source normalization and mass correction are
\begin{equation}
\mbox{$\displaystyle \begin{aligned}
U_0={}&b_d\left[R^{2-d}+\frac{A_d^2}{3}R^{6-3d}+\cdots\right],\\
U-U_0={}&\mu^2b_d\biggl[\frac{R^{4-d}}{2(4-d)}\\
&+\frac{4A_dR^{6-2d}}{(d-2)(6-2d)(4-d)}+\cdots\biggr]+\cdots.
\end{aligned}$}
\end{equation}
Using \mbox{$\displaystyle \nabla_d^2R^s=s(s+d-2)R^{s-2}$} in Eq.~(\ref{eq:unit-source-dr-operator}) gives the coefficient multiplying \mbox{$\displaystyle A_d^2R^{8-3d}$} in the mass correction after factoring out \mbox{$\displaystyle \mu^2b_d$}
\begin{equation}
\mbox{$\displaystyle \begin{aligned}
&\frac{(d+2)(4-d)/(d-2)^2+1/3+(3-d)}{(8-3d)(6-2d)}\\
&\qquad=\frac{8}{3(d-3)}-\frac{31}{2}+\mathcal O(d-3).
\end{aligned}$}
\end{equation}
The term \mbox{$\displaystyle 3-d$} comes from the kinetic curvature correction. Its product with the pole contributes to the finite result and must be kept before taking $d\to3$.

The dimensional prefactors and radial powers also contribute finite terms when multiplied by the pole. Their expansions are
\begin{equation}
\mbox{$\displaystyle \begin{aligned}
b_dR^{2-d}={}&\frac1R\biggl\{1+(d-3)\biggl[\log\frac\ell R\\
&-\frac{\gamma_{\rm E}+\log(4\pi)}2\biggr]+\cdots\biggr\},\\
A_dR^{2-d}={}&\frac M{2R}\biggl\{1+(d-3)\biggl[\log\frac\ell R\\
&-\frac{\gamma_{\rm E}+\log(4\pi)}2+\frac12\biggr]+\cdots\biggr\}.
\end{aligned}$}
\end{equation}
Before subtraction, the second-order monopole term is therefore
\begin{equation}
\mbox{$\displaystyle \begin{aligned}
&\mu^2b_dA_d^2R^{8-3d}\left[\frac8{3(d-3)}-\frac{31}2+\cdots\right]\\
&\quad=\frac{\mu^2M^2}{R}\biggl[\frac2{3(d-3)}+2\log\frac\ell R\\
&\hspace{25mm}-\gamma_{\rm E}-\log(4\pi)-\frac{77}{24}+\cdots\biggr].
\end{aligned}$}
\end{equation}

Subtracting only the poles relates the bare source coefficient to the worldline coupling through
\begin{equation}
\mbox{$\displaystyle Q_{\rm b}=Q(\ell)
\left[1-\frac{2\mu^2M^2}{3(d-3)}\right]+\cdots.$}
\end{equation}
The counterterm multiplies the unexpanded leading field \mbox{$\displaystyle b_dR^{2-d}$}. Its finite contribution is
\begin{equation}
\mbox{$\displaystyle \frac{\mu^2M^2}{R}\left[-\frac23\log\frac\ell R+\frac{\gamma_{\rm E}+\log(4\pi)}3\right].$}
\end{equation}
After subtraction, the displayed monopole sector of \mbox{$\displaystyle U$} is finite in the same convention as the coupling. Expanding both factors before taking the finite part yields
\begin{equation}
\label{eq:unit-source-finite-monopole}
\mbox{$\displaystyle \begin{aligned}
\left.U\right|_{\mu^2M^2/R}={}&\frac{\mu^2M^2}{R}\biggl[-\frac43\log\frac R\ell\\
&-\frac23\bigl(\gamma_{\rm E}+\log(4\pi)\bigr)-\frac{77}{24}\biggr].
\end{aligned}$}
\end{equation}
Linear derivative operators proportional to $\tilde\Box\phi$ are included in this coupling by using the homogeneous external-field equation $\tilde\Box\phi=\mu^2\phi$~\cite{Diedrichs2024MassivePhase}.

Transforming the complete BH near solution to the same radius gives its monopole sector
\begin{equation}
\label{eq:bh-isotropic-monopole}
\mbox{$\displaystyle \left.\frac{\phi_{\rm BH}}{2\beta_0/M}\right|_{1/R}=\frac1R\biggl\{1+\mu^2M^2\biggl[\frac{19}{24}-\frac43\log\frac R{r_h}\biggr]\biggr\}+\cdots.$}
\end{equation}
The constant is $19/24=1/8+2-4/3$, with contributions from the growing radial term, the logarithm, and the areal monopole, respectively. The common weak-zone GB particular field starts with $-4\beta_0M^2/R^4$ and $-2\beta_0\mu^2M^2/R^2$. These terms have different radial powers from the monopole and are kept separate in the comparison. The global forced tail remains part of the complete field. Comparing Eqs.~(\ref{eq:unit-source-finite-monopole}) and (\ref{eq:bh-isotropic-monopole}), the finite constants add to $19/24+77/24=4$, giving the conversion in Eq.~(\ref{eq:static-worldline-conversion}).

For a check using the Green identity, we use a regular homogeneous solution whose normalization is fixed by its Wronskian with the unit-source solution. In isotropic coordinates, the normalization is
\begin{equation}
\mbox{$\displaystyle \left(R^2-\frac{M^2}{4}\right)\left(H\frac{\mathrm{d}U}{\mathrm{d}R}-U\frac{\mathrm{d}H}{\mathrm{d}R}\right)=-1,$}
\end{equation}
through the retained order. The source solution and the regular homogeneous solution use the same subtraction scheme and are normalized to preserve the Wronskian fixed by \mbox{$\displaystyle U\sim1/R$} and \mbox{$\displaystyle H\sim1$}. The weak-field regular solution is
\begin{equation}
\mbox{$\displaystyle \begin{aligned}
H={}&1+\mu^2\biggl\{\frac{R^2}{6}+MR+M^2\biggl[\frac43\log\frac R\ell\\
&+\frac23\bigl(\gamma_{\rm E}+\log(4\pi)\bigr)+\frac{17}{36}\biggr]+\cdots\biggr\}
+\cdots.
\end{aligned}$}
\end{equation}
The finite constant \mbox{$\displaystyle 17/36$} is fixed by the Wronskian with the unit-source field. Its sum with \mbox{$\displaystyle -77/24$} is \mbox{$\displaystyle -197/72$}, which cancels the remaining constant at order \mbox{$\displaystyle \mu^2M^2$} in that Wronskian. The solution regular at the horizon is
\begin{equation}
\label{eq:regular-source-test-field}
\mbox{$\displaystyle \begin{aligned}
H={}&1+\epsilon^2h_2+\cdots,\\
h_2={}&\frac{z^2}{6}+\frac z3+\frac{\log z}{3}-\frac19\\
&+\frac{\gamma_{\rm E}+\log(4\pi)}{6}
+\frac13\log\frac{r_h}{\ell}.
\end{aligned}$}
\end{equation}
Applying the Green identity to the GB-sourced field order by order gives
\begin{equation}
\mbox{$\displaystyle \begin{aligned}
\frac{r_hQ(\ell)}{\beta_0}
={}&12\int_1^\infty\frac{1+\epsilon^2h_2(z)}{z^4}\,\mathrm{d}z+\cdots\\
={}&4+\epsilon^2\biggl[4+\frac23\bigl(\gamma_{\rm E}+\log(4\pi)\bigr)\\
&\hspace{24mm}+\frac43\log\frac{r_h}{\ell}\biggr]+\cdots.
\end{aligned}$}
\end{equation}
The integral acts on the mass-expansion coefficients, not on an exact growing massive solution. It reproduces the direct field matching. The finite constant in the worldline coefficient belongs to the stated subtraction convention. Through this order the subtraction-scale dependence cancels in the monopole sector of the source field, including its logarithmic terms
\begin{equation}
\mbox{$\displaystyle \frac{\mathrm{d}}{\mathrm{d}\log\ell}
\left[Q(\ell)U(R;\ell)\right]_{1/R}=0.$}
\end{equation}
The analytic quadratic coefficient was checked independently by numerically reconstructing the perturbative $\mathcal O(\epsilon^2)$ radial problem. The regular homogeneous constant is fixed by $\psi_2(1)=0$, and the numerical reconstruction is sampled on a logarithmic radial grid over $25\leq z\leq1800$. The $1/z$ coefficient is kept free in the large-$z$ fit, providing an independent numerical test of the analytic coefficient. At asymptotic order $N$, the fitting basis contains $z$, $\log z$, $1$, $(\log z)/z$, and $1/z$, together with the pairs $\{(\log z)/z^k,1/z^k\}_{k=2}^{N}$. Table~\ref{tab:quadratic-reconstruction-convergence} shows the convergence for $N=2$, $3$, and $4$ over three large-radius fitting windows.

\begin{table}[!ht]
\centering
\caption{Convergence of the independent numerical reconstruction of the quadratic $1/z$ coefficient in the static near-zone field.}
\label{tab:quadratic-reconstruction-convergence}
\begin{tabular*}{\columnwidth}{@{\extracolsep{\fill}}cccc@{}}
\toprule
Fit window & $N=2$ & $N=3$ & $N=4$ \\
\midrule
$80\leq z\leq800$
& $-1.3380350$
& $-1.3333801$
& $-1.3333337$ \\
$100\leq z\leq1000$
& $-1.3367598$
& $-1.3333612$
& $-1.3333335$ \\
$150\leq z\leq1500$
& $-1.3351653$
& $-1.3333434$
& $-1.3333334$
\\
\bottomrule
\end{tabular*}
\end{table}

The largest change between the $N=3$ and $N=4$ fits in Table~\ref{tab:quadratic-reconstruction-convergence} is approximately $4.6\times10^{-5}$. The median of the three $N=4$ window values is $c_2^{\rm num}=-1.33333$. This independent reconstruction is consistent with the analytic coefficient $c_2=-4/3$. The finite-$\epsilon$ BVP tests the radial construction and far-basis extraction, whereas the perturbative reconstruction provides the independent numerical check of the quadratic local-matching coefficient.

For the representative finite-mass checks, collocation uses a logarithmic radius and a tolerance of $2\times10^{-9}$. The outer decaying condition converts the radial logarithmic derivative to that coordinate before imposing the Robin relation. Shooting and collocation are compared with the same forced subtraction and tenth-order homogeneous basis over $10\leq\epsilon z\leq16$. At these three masses, a finite-radius Green projection of the complete numerical field also checks the implementation with the analytically fixed homogeneous normalization.

The projection uses the full numerical field and a numerically integrated horizon-regular homogeneous solution, normalized through second order as in Eq.~(\ref{eq:regular-source-test-field}). With primes denoting derivatives with respect to \mbox{$\displaystyle z$}, the finite-radius form of the Green identity is
\begin{equation}
\mbox{$\displaystyle \begin{aligned}
\frac{r_hQ(\ell)}{\beta_0}={}&-z(z-1)\left(H\psi'-H'\psi\right)\\
&+12\int_z^\infty\frac{1+\epsilon^2h_2(t)}{t^4}\,\mathrm{d}t+\cdots.
\end{aligned}$}
\end{equation}
The boundary term retains the complete numerical field, including the distributed GB source. The integral adds the exterior source contribution expanded through second order. At \mbox{$\displaystyle \ell=r_h$} and \mbox{$\displaystyle z=5,8,10$}, Table~\ref{tab:worldline-projection} compares the coefficient extracted from this sum with the analytic value \mbox{$\displaystyle 6.072159941\ldots$} in the expansion of \mbox{$\displaystyle r_hQ/\beta_0$}. The coefficient is extracted by subtracting \mbox{$\displaystyle 4$} and dividing by \mbox{$\displaystyle \epsilon^2$}.

\begin{table}[!ht]
\centering
\caption{Finite-radius Green-identity consistency check at \mbox{$\displaystyle \ell=r_h$} and \mbox{$\displaystyle z=5,8,10$}. The homogeneous normalization is fixed analytically. The last column is the projected dimensionless source coefficient minus its second-order analytic expression, divided by \mbox{$\displaystyle \epsilon^4$}.}
\label{tab:worldline-projection}
\begin{tabular*}{\columnwidth}{@{\extracolsep{\fill}}cccc@{}}
\toprule
\mbox{$\displaystyle \epsilon$} & \mbox{$\displaystyle z$} & Quadratic coefficient & Residual/\mbox{$\displaystyle \epsilon^4$} \\
\midrule
\mbox{$\displaystyle 0.01$} & \mbox{$\displaystyle 5$} & \mbox{$\displaystyle 6.0724226901$} & \mbox{$\displaystyle 2.62749$} \\
\mbox{$\displaystyle 0.01$} & \mbox{$\displaystyle 8$} & \mbox{$\displaystyle 6.0724965732$} & \mbox{$\displaystyle 3.36632$} \\
\mbox{$\displaystyle 0.01$} & \mbox{$\displaystyle 10$} & \mbox{$\displaystyle 6.0725345214$} & \mbox{$\displaystyle 3.74580$} \\
\addlinespace[2pt]
\mbox{$\displaystyle 0.02$} & \mbox{$\displaystyle 5$} & \mbox{$\displaystyle 6.0732148610$} & \mbox{$\displaystyle 2.63730$} \\
\mbox{$\displaystyle 0.02$} & \mbox{$\displaystyle 8$} & \mbox{$\displaystyle 6.0735100052$} & \mbox{$\displaystyle 3.37516$} \\
\mbox{$\displaystyle 0.02$} & \mbox{$\displaystyle 10$} & \mbox{$\displaystyle 6.0736620725$} & \mbox{$\displaystyle 3.75533$} \\
\addlinespace[2pt]
\mbox{$\displaystyle 0.05$} & \mbox{$\displaystyle 5$} & \mbox{$\displaystyle 6.0787518943$} & \mbox{$\displaystyle 2.63678$} \\
\mbox{$\displaystyle 0.05$} & \mbox{$\displaystyle 8$} & \mbox{$\displaystyle 6.0806056535$} & \mbox{$\displaystyle 3.37828$} \\
\mbox{$\displaystyle 0.05$} & \mbox{$\displaystyle 10$} & \mbox{$\displaystyle 6.0815593631$} & \mbox{$\displaystyle 3.75977$} \\
\bottomrule
\end{tabular*}
\end{table}

The projected quadratic coefficient approaches the analytic value as \mbox{$\displaystyle \epsilon$} decreases at each of the three radii. The spread between the finite-radius estimates also decreases with \mbox{$\displaystyle \epsilon$}. The scaled residual is nearly constant at each fixed radius, supporting the second-order implementation without determining the full higher-order remainder. Since the projection is unchanged by adding a multiple of \mbox{$\displaystyle H$} to \mbox{$\displaystyle \psi$}, it does not independently determine the finite normalization constant or select the global decaying solution.

\setcounter{equation}{0}
\section{Worldline normalization and leading scalar radiation}
\label{app:worldline-radiation}

This appendix derives the normalization of the worldline source, Yukawa exchange, and leading scalar-radiation terms used in Secs.~\ref{sec:conservative-dynamics} and \ref{sec:scalar-waveform}. The binary effective description begins with~\cite{Julie2019Dynamics,Kuntz2019EFT}
\begin{equation}
S_{\rm pp}=-\sum_A\int m_A(\phi;\mu)\,\mathrm{d}\tilde s_A.
\end{equation}
Expanding the mass function about a vanishing external scalar field gives
\begin{equation}
\begin{aligned}
m_A(\phi;\mu)&=m_A-\frac{Q_A}{4}\phi+\mathcal O(\phi^2),\\
Q_A&=-4\,\partial_\phi m_A(\phi;\mu)\big|_{\phi=0}.
\end{aligned}
\end{equation}
The mass $m_A\equiv m_A(0;\mu)$ is the background component mass, and $\mathrm{d}\tilde s_A$ is the Einstein-frame worldline line element. In the weak zone relevant to the retained point-source approximation, $\tilde g_{\mu\nu}$ reduces to the Minkowski metric $\eta_{\mu\nu}$ and $\tilde\Box$ reduces to the flat d'Alembertian $\Box_\eta$. The linear term then gives
\begin{equation}
S_Q=\frac{1}{4}\int\mathrm{d}^4x\,\rho_Q\phi,\qquad
\rho_Q=\sum_A Q_A\delta^{(3)}(\mathbf x-\mathbf z_A).
\end{equation}
The coefficient $1/4$ follows from the scalar kinetic normalization and the chosen point-source normalization, while the physical-frame matter coupling remains separately controlled by $\alpha_0$.

With the canonical scalar normalization of Eq.~(\ref{eq:effective-action}), variation in this weak-zone limit yields
\begin{equation}
\label{eq:appendix-weak-zone-scalar-equation}
(\Box_\eta-\mu^2)\phi=-4\pi\rho_Q.
\end{equation}
The curvature-sourced strong-field response has already been encoded in $Q_A$ through the isolated-BH matching. A distributed GB source also exists in the binary weak zone, but the scaling in Eq.~(\ref{eq:weak-zone-gb-suppression}) suppresses it by $v^6$, or 3PN, relative to the matched monopole field. Equation~(\ref{eq:appendix-weak-zone-scalar-equation}) therefore keeps only the effective monopole source $\rho_Q$ at the order retained here.
Comparing the $1/r$ terms of the black-hole and point-source fields in the same coordinates and scalar normalization fixes the leading monopole normalization. This appendix uses the leading matched charges, while the second-order static conversion is derived in \ref{app:scalar-matching}.

For a field point $\mathbf x$, let \mbox{$\displaystyle r_A=|\mathbf x-\mathbf z_A|$} denote its distance from body $A$. At leading point-source order, the scalar field in the part of the binary EFT buffer region where \mbox{$\displaystyle \mu r_A\ll1$} satisfies
\begin{equation}
\mbox{$\displaystyle \phi_A\simeq\frac{Q_A}{r_A},$}
\end{equation}
which is the monopole reproduced by the worldline source in the same scalar normalization and matching prescription.
For a static source, the Yukawa Green function satisfies
\begin{equation}
(\nabla^2-\mu^2)\frac{\mathrm{e}^{-\mu|\mathbf x-\mathbf x'|}}
{|\mathbf x-\mathbf x'|}
=-4\pi\delta^{(3)}(\mathbf x-\mathbf x').
\end{equation}
The field generated by body $A$ is therefore $\phi_A=Q_A\mathrm{e}^{-\mu r_A}/r_A$. This reproduces the matched monopole normalization used in the binary weak zone.

The on-shell scalar action makes the exchange sign explicit. After integration by parts and use of Eq.~(\ref{eq:appendix-weak-zone-scalar-equation}), the scalar kinetic, mass, and source terms reduce to
\begin{equation}
\left.S_{\phi,Q}\right|_{\rm on\text{-}shell}
=\frac{1}{8}\int\mathrm{d}^4x\,\rho_Q\phi.
\end{equation}
For two distinct sources, the two equal cross terms remain after omitting self-energy. They give
\begin{equation}
L_{\rm cross}=\frac{Q_1Q_2}{4r}\mathrm{e}^{-\mu r},
\qquad
U_\phi=-L_{\rm cross}.
\end{equation}
This establishes the factor and the attractive sign in Eq.~(\ref{eq:yukawa-potential}).

The scalar source multipoles are defined by
\begin{equation}
D_Q^i=\sum_AQ_Az_A^i,\qquad
I_Q^{ij}=\sum_AQ_Az_A^{\langle ij\rangle}.
\end{equation}
In the center-of-mass frame, $\mathbf z_1=(m_2/m)\mathbf r$ and $\mathbf z_2=-(m_1/m)\mathbf r$. Substitution of $Q_A=2m_Aq_A$ gives
\begin{equation}
\label{eq:appendix-com-multipoles}
D_Q^i=2\eta m\Delta q\,r^i,\qquad
I_Q^{ij}=2\eta m\bar q\,r^{\langle ij\rangle}.
\end{equation}
At the retained linear worldline-response order, the total scalar monopole $\sum_AQ_A$ is time independent for the quasicircular system and has no positive-frequency radiative component. The leading radiative source multipoles are therefore the dipole and quadrupole in Eq.~(\ref{eq:appendix-com-multipoles}).

For a radiative harmonic, let $\omega_n=n\Omega$ and $k_n^2=\omega_n^2-\mu^2$. A propagating mode requires $\omega_n>\mu$, for which $k_n$ is real, while below threshold the radial mode is evanescent. Let $R$ denote the wave-zone radial distance and $\mathbf N$ the radiation-direction unit vector. For the complex Fourier components used in this appendix, we take the time dependence as $\mathrm{e}^{-i\omega_n t}$, so the outgoing radial factor is $\mathrm{e}^{+ik_nR}$. The quantities $D_{Q,\omega_n}^i$ and $I_{Q,\omega_n}^{ij}$ are the Fourier components of $D_Q^i$ and $I_Q^{ij}$ at that harmonic. Its leading multipole form is
\begin{equation}
\label{eq:appendix-far-zone-multipole-field}
\mbox{$\displaystyle \phi_{\omega_n}(R,\mathbf N)=\frac{\mathrm{e}^{ik_nR}}{R}\biggl[-ik_nN_iD_{Q,\omega_n}^i-\frac{k_n^2}{2}N_iN_jI_{Q,\omega_n}^{ij}+\cdots\biggr].$}
\end{equation}
In the weak wave zone, the scalar part of Eq.~(\ref{eq:effective-action}) gives the canonical stress-energy tensor
\begin{equation}
\begin{aligned}
T_{\phi}^{\mu\nu}={}&\frac{1}{16\pi}\biggl[\partial^\mu\phi\,\partial^\nu\phi\\
&-\frac{1}{2}\eta^{\mu\nu}\left(\partial_\alpha\phi\,\partial^\alpha\phi+\mu^2\phi^2\right)\biggr].
\end{aligned}
\end{equation}
Let $\mathcal B(\mathbf N)$ denote the dipole-plus-quadrupole angular amplitude inside the square brackets in Eq.~(\ref{eq:appendix-far-zone-multipole-field}). The time-averaged radial flux density is
\begin{equation}
\left\langle T_\phi^{0R}\right\rangle
=\frac{\omega_n k_n}{32\pi R^2}\left|\mathcal B(\mathbf N)\right|^2,
\end{equation}
where outer brackets denote a time average and $T_\phi^{0R}$ is the radial energy-flux component. The factor $1/(32\pi)$ follows from the scalar normalization in Eq.~(\ref{eq:effective-action}) together with the time average of the real monochromatic field.

The required isotropic angular integrals are
\begin{equation}
\begin{aligned}
\int N_iN_j\,\mathrm{d}\Omega
&=\frac{4\pi}{3}\delta_{ij},\\
\int N_iN_jN_kN_l\,\mathrm{d}\Omega
&=\frac{4\pi}{15}
\left(
\delta_{ij}\delta_{kl}
+\delta_{ik}\delta_{jl}
+\delta_{il}\delta_{jk}
\right).
\end{aligned}
\end{equation}
The dipole contribution follows from the first angular integral. For the quadrupole contribution, the STF property removes the trace contraction from the four-index integral and leaves the two equal contractions, giving
\begin{equation}
\label{eq:appendix-general-scalar-fluxes}
\mbox{$\displaystyle \mathcal F_{\phi}^{\rm dip}=\frac{\omega_1 k_1^3}{24}D_{Q,\omega_1}^{i}D_{Q,\omega_1}^{i*},\qquad\mathcal F_{\phi}^{\rm quad}=\frac{\omega_2 k_2^5}{240}I_{Q,\omega_2}^{ij}I_{Q,\omega_2}^{ij*}.$}
\end{equation}

For a circular orbit at a fixed adiabatic frequency, take $\mathbf r=r(\cos\Omega t,\sin\Omega t,0)$. The Fourier amplitudes implied by Eq.~(\ref{eq:appendix-com-multipoles}) satisfy
\begin{equation}
\begin{aligned}
D_{Q,\omega_1}^{i}D_{Q,\omega_1}^{i*}
&=
8\eta^2m^2(\Delta q)^2r^2,\\
I_{Q,\omega_2}^{ij}I_{Q,\omega_2}^{ij*}
&=
4\eta^2m^2\bar q^2r^4.
\end{aligned}
\end{equation}
The scalar fluxes are already quadratic in the matched charges, so the leading circular relations are sufficient in this step. Inserting the $\mathcal O(\kappa)$ correction to the orbit would contribute beyond the retained $\mathcal O(q_A^2)$ scalar-radiation order. Using $r=m/x$ and $\Omega=x^{3/2}/m$, with $k_1=\Omega v_1$ and $k_2=2\Omega v_2$, Eq.~(\ref{eq:appendix-general-scalar-fluxes}) reduces to
\begin{equation}
\begin{aligned}
\mathcal F_\phi^{\rm dip}
&=
\frac{1}{3}\eta^2(\Delta q)^2x^4v_1^3
\Theta(\Omega-\mu),\\
\mathcal F_\phi^{\rm quad}
&=
\frac{16}{15}\eta^2\bar q^2x^5v_2^5
\Theta(2\Omega-\mu).
\end{aligned}
\end{equation}
The powers $v_1^3$ and $v_2^5$ follow from the factors $\omega_1k_1^3$ and $\omega_2k_2^5$, respectively. The step functions restrict these expressions to propagating modes above the corresponding massive thresholds. These results reproduce the scalar terms in Eq.~(\ref{eq:leading-fluxes}) and agree with the standard massive-scalar radiation structure~\cite{Diedrichs2024MassivePhase,Liu2020MassiveBDWaveforms}.

The same far-zone multipole field also fixes the source amplitudes used in Eq.~(\ref{eq:scalar-time-domain-harmonics}). With $\mathbf N=(\sin\iota,0,\cos\iota)$, the dipole projection carries the $n=1$ angular factor $\sin\iota$, while the radiative quadrupole carries the $n=2$ factor $\sin^2\iota$. Using $k_1r=x^{1/2}v_1$ and $k_2^2r^2/4=xv_2^2$, and taking the real part with the convention of Sec.~\ref{sec:scalar-waveform} reproduces the amplitudes in Eq.~(\ref{eq:scalar-time-domain-harmonics}).

\bibliographystyle{spphys}
\bibliography{references}

\end{document}